\documentclass{emulateapj}
\usepackage{aas_macros}
\usepackage{epsfig}
\usepackage{multirow}
\usepackage{longtable}
\begin{document}

\title{Carbon Stars in the Satellites and Halo of M31}

\author{Katherine Hamren\altaffilmark{1,2}, Rachael L. Beaton\altaffilmark{3}, Puragra Guhathakurta\altaffilmark{1}, Karoline M. Gilbert\altaffilmark{4,5}, Erik J. Tollerud\altaffilmark{4}, Martha L. Boyer\altaffilmark{6}, Constance M. Rockosi\altaffilmark{1}, Graeme H. Smith\altaffilmark{1}, Steven R. Majewski\altaffilmark{7}, Kirsten Howley\altaffilmark{8}}

\altaffiltext{1}{Department of Astronomy and Astrophysics, University of California Santa Cruz, 1156 High Street, Santa Cruz, CA  95064, USA}
\altaffiltext{2}{{\tt khamren@ucolick.org}}
\altaffiltext{3}{ The Observatories of the Carnegie Institutions for Science, 813 Santa Barbara Street, Pasadena, CA 91101, USA}
\altaffiltext{4}{Space Telescope Science Institute, Baltimore, MD 21218, USA}
\altaffiltext{5}{Center for Astrophysical Sciences, Johns Hopkins University, Baltimore, MD, 21218, USA}
\altaffiltext{6}{Observational Cosmology Lab, Code 665, NASA Goddard Space Flight Center, Greenbelt, MD 20771, USA}
\altaffiltext{7}{Department of Astronomy, University of Virginia, Charlottesville, VA 22904, USA}
\altaffiltext{8}{Lawrence Livermore National Laboratory, P.O. Box 808, Livermore, CA 94551, USA}

\date{}

\begin{abstract}
We spectroscopically identify a sample of carbon stars in the satellites and halo of M31 using moderate-resolution optical spectroscopy from the Spectroscopic and Photometric Landscape of Andromeda's Stellar Halo survey. We present the photometric properties of our sample of 41 stars, including their brightness with respect to the tip of the red giant branch (TRGB) and their distributions in various color-color spaces. This analysis reveals a bluer population of carbon stars fainter than the TRGB and a redder population of carbon stars brighter than the TRGB. We then apply principal component analysis to determine the sample's eigenspectra and eigencoefficients. Correlating the eigencoefficients with various observable properties reveals the spectral features that trace effective temperature and metallicity. Putting the spectroscopic and photometric information together, we find the carbon stars in the satellites and halo of M31 to be minimally impacted by dust and internal dynamics. We also find that while there is evidence to suggest that the sub-TRGB stars are extrinsic in origin, it is also possible that they are are particularly faint members of the asymptotic giant branch. 

\end{abstract}


\maketitle

\section{Introduction}

Carbon stars are nominally defined as stars with more free carbon than free oxygen in their atmospheres. This excess carbon builds up via the third dredge-up (TDU) process in thermally pulsating asymptotic giant branch (TP-AGB) stars. The TP-AGB stage is characterized by unstable double shell burning. TDU occurs when the He-burning shell around the AGB star's inert C+O core ignites and extinguishes the outer H-burning shell. This allows the star's outer convective envelope to penetrate the intershell region and ``dredge" $^{12}$C up to the surface. Over time and successive dredge ups, the ratio (C/O) of $^{12}$C to $^{16}$O  increases, eventually exceeding unity. The stars in which C/O$> 1$ are carbon stars (C-stars). 

However, carbon stars have been observed at luminosities below the (mass) limit necessary for TDU. This ever growing population of faint carbon stars includes CH stars, dwarf carbon (dC) stars, and carbon enhanced metal poor (CEMP) stars. These faint stars are theorized to have received their carbon via accretion from a carbon-rich AGB companion rather than internal processes \citep[][]{deKool1995, Frantsman1997, Izzard2004}. Faint carbon stars are thus often termed ``extrinsic", in contrast to their ``intrinsic" TP-AGB counterparts.

Carbon stars make very unique tracers of particular stellar populations. Carbon-rich TP-AGB stars are easily identified photometrically or spectroscopically, and are more difficult to confuse with other tracer populations (e.g., as is the case between blue stragglers and blue horizontal branch stars). As a result, they have been used to map morphological structure, kinematical structure, mean age and metallicity of various hosts \citep[e.g.,][]{Rowe2005, Battinelli2005, Demers2007, Cioni2008, Huxor2015}. These stars also trace intermediate-age populations, and so have been used to constrain the star formation histories of various Local Group objects \citep[][and references therein]{Grebel2007}. Extrinsic carbon stars provide information about earlier generations of AGB stars as well as the binary systems in which they are found.

Both intrinsic and extrinsic carbon stars have been identified throughout the Local Group. There have been dedicated carbon star surveys, or AGB surveys for which carbon stars share priority with oxygen-rich M-stars, using photometry in the optical \citep[e.g.,][and subsequent papers in these series]{Albert2000, Nowotny2001}, near-infrared \citep[e.g.,][]{Cioni2005, Whitelock2006, Battinelli2007} and mid-infrared \citep[e.g.,][]{Blum2006, Boyer2011a, Woods2011}. In addition, non-dedicated surveys have been mined for carbon stars, AGB or otherwise \citep[e.g.,][]{Margon2002,Green2013, Hamren2015}. In many nearby dwarf galaxies, and various fields in the Milky Way (MW) halo, these surveys have enabled detailed abundance studies \citep[e.g.,][]{Abia1993, Abia2002}. However, the more distant satellite galaxies associated with M31 are fainter, and often difficult to distinguish from the MW foreground and the M31 halo. As a result, studies in these objects have largely been limited to C-star identification and the ratio (C/M) of carbon- to oxygen-rich TP-AGB stars. 

Recent advances have made these previously under-studied regions ripe for further attention. Large-scale surveys like the Spectroscopic Landscape of Andromeda's Stellar Halo \citep[SPLASH - ][]{Guhathakurta2005, Guhathakurta2006, Tollerud2012, Gilbert2012, Gilbert2014, Dorman2012, Dorman2015} and the Pan-Andromeda Archaeological Survey \citep[PAndAS - ][]{McConnachie2009} have produced a wealth of spectroscopic and photometric data that can be used to study carbon stars in a uniform way. In addition, the M31 satellites have been shown via thorough characterization of their kinematical properties to be low-dispersion systems \citep{Tollerud2012, Ho2012, Collins2013, Tollerud2013}. This improves the veracity of kinematical determination of satellite membership. Finally, several M31 satellites now have star formation histories (SFHs) derived from deep \textit{Hubble Space Telescope} (HST) images \citep{Weisz2014, Geha2015} and identification of carbon stars can be put into a far broader context.

In this work we use photometric and spectroscopic data from SPLASH to study carbon stars in the satellites and halo of M31. Section~\ref{data} describes our dataset, including a summary of the observations, the photometric transformations and synthetic photometry used to homogenize the sample, and criteria for determining membership. Section~\ref{cstars} discusses the carbon stars themselves; identification, location, and the impact of the SPLASH selection function on the final sample. Section~\ref{photometry} looks at the photometric properties of the carbon stars, while Section~\ref{spectroscopy} looks at their spectroscopic properties. We discuss the implications of our findings in Section~\ref{discussion}.

\section{Data}\label{data}

\begin{figure*}[t!]
\centering
\includegraphics[width=7in]{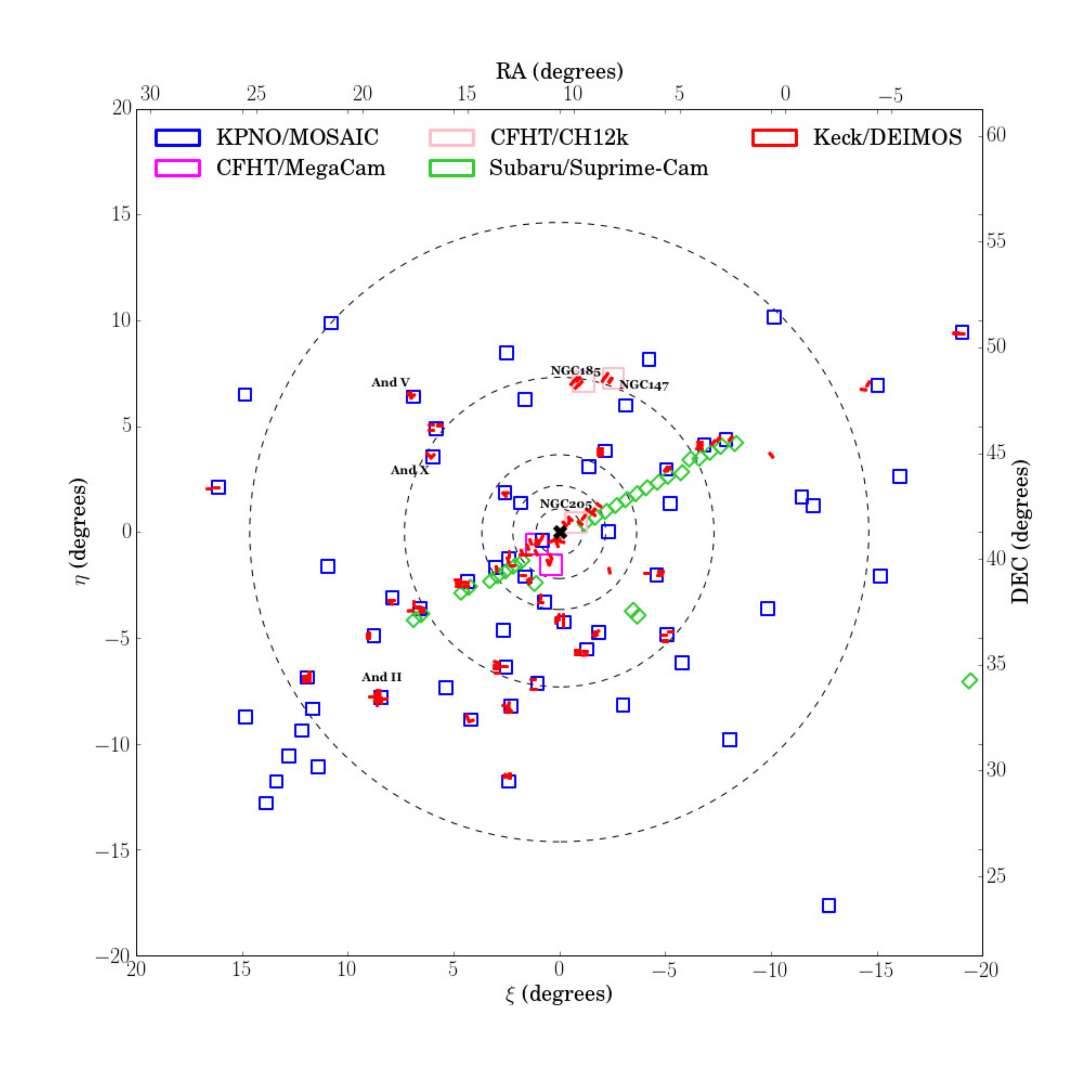}
\caption[SPLASH Survey Map]{SPLASH survey map (including M32 but excluding the bright disk of M31). Footprints of the images used for spectroscopic target selection are shown to scale as blue (KPNO/MOSAIC), magenta (CFHT/MegaCam), pink (CFHT/CFH12k) or green (Subaru/Suprime-Cam) rectangles. The footprints of the DEIMOS masks are shown to scale as red rectangles. The center of M31 itself is marked as a bold black $\times$, and several of the more prominent satellite galaxies are labeled. The dashed circles represent projected radii of 15, 30, 50, 100 and 200kpc. }
\label{fig:survey_map}
\end{figure*}

\begin{deluxetable*}{lccccccccc}
\tablewidth{0pt}
\tablecaption{Properties of M31 Satellites Observed by SPLASH}
\tablehead{Satellite & RA\tablenotemark{a} & DEC\tablenotemark{a} & [Fe/H]\tablenotemark{a} & $M_{\rm V}$\tablenotemark{a} & $(m-M)_0$\tablenotemark{a} & $v_{\rm sys}$ (km s$^{-1}$)\tablenotemark{b} & $\sigma$ (km s$^{-1}$)\tablenotemark{b} &  [$\alpha$/Fe]\tablenotemark{h} & $I_{\rm TRGB}$\tablenotemark{i}} \\
\startdata
And I & 00:45:39.8 & +38:02:28 & $-1.45\pm0.04$ & $-11.7\pm0.1$ & $24.36\pm0.07$ & $-376 \pm 2.2$& $10.2 \pm 1.9$ & $0.278\pm 0.164$ & 20.36\\
And II & 01:16:29.8 & +33:25:09 & $-1.64 \pm 0.04$ & $-12.4 \pm 0.2$ & $24.07 \pm 0.06$ & $-192.4 \pm 0.5$\tablenotemark{c} & $7.8 \pm 1.1$\tablenotemark{c} & $0.033\pm 0.09$ & 20.10\\
And III & 00:35:33.8 & +36:29:52 & $-1.78\pm 0.04$ & $ -10.0\pm0.3$ & $24.37\pm 0.07$&  $-344.3 \pm 1.7$ & $0.3 \pm 1.4$ & $0.205 \pm 0.16$ & 20.40\\
And V & 01:10:17.1 & +47:37:41 & $-1.6 \pm 0.3$ & $-9.1 \pm 0.2$ & $24.44 \pm 0.08$ & $-397.3 \pm 1.5$ & $10.5 \pm 1.1$ & $0.119 \pm 0.09$ & 20.54 \\
And VII & 23:26:31.7 &+50:40:33 & $ -1.40\pm0.30$ & $-12.6\pm0.3$ & $24.41  \pm0.10$ & $ -307.2\pm1.3$ & $13.0\pm1.0$ & $0.296\pm 0.09$ & 20.59\\
And IX & 00:52:53.0 & +43:11:45 & $-2.2 \pm 0.2$ & $-8.1 \pm 1.1$ & $24.42 \pm 0.07$ & $-209.4 \pm 2.5$ & $10.9 \pm 2.0$& ...& 20.59\\
And X & 01:06:33.7 & +44:48:16 & $-1.93 \pm 0.11$ & $-7.6 \pm 1.0$ & $24.23 \pm 0.21$ & $-164.1 \pm 1.7$ & $6.3 \pm 1.4$ & $0.505\pm 0.24$ & 20.33\\
And XI & 00:46:20.0 & +33:48:05 & $-2.00\pm0.20$ & $ -6.9\pm1.3$ & $24.40^{+0.20}_ {-0.50}$ & $-427.5 \pm 3.5$\tablenotemark{d} & $7.6^{+4.0}_{-2.8}$\tablenotemark{d} &... & 20.51\\
And XII & 00:47:27.0 &+34:22:29 & $-2.10\pm0.20$ & $ -6.4\pm1.2$ &  $24.70\pm0.30$ &  $-557.1\pm 1.7$\tablenotemark{d} & $0.0^{+4.0}$\tablenotemark{d} & ...& 20.88\\
And XIII & 00:51:51.0 & +33:00:16 & $-1.90\pm0.20$ & $-6.7\pm1.3$ & $24.80^{+0.10}_{-0.40}$ &  $-185.4 \pm 2.4$ & $5.8 \pm 2.0$ &... & 20.89\\
And XIV & 00:51:35.0 & +29:41:49 & $-2.26\pm0.05$ & $-8.4\pm0.6$ & $24.33\pm0.33$ & $-480.6 \pm 1.2$ & $5.3\pm 1.0$ & ...& 20.51\\
And XV & 01:14:18.7 & +38:07:03 & $-1.80\pm0.20$ & $-9.4 \pm0.4$ &  $24.00\pm0.20$ & $-323.0 \pm 1.4$ & $4.0\pm1.4$ & ...& 20.02 \\
And XVI & 00:59:29.8 & +32:22:36 & $-2.10\pm0.20$ & $-9.2\pm0.4$ & $23.60 \pm 0.20$ &  $-367.3\pm2.8$ & $3.8\pm2.9$ &... & 19.72\\
And XVIII & 00:02:14.5 & +45:05:20 & $-1.8 \pm 0.1$ & $-9.7$ & $25.66 \pm 0.13$ & $-332.1\pm2.7$ & $9.7\pm2.3$& ...& 21.76 \\
And XXI & 23:54:47.7 & +42:28:15 &$-1.80\pm0.20$ & $ -9.9\pm0.6$ & $24.67\pm0.13$ &  $-361.4\pm 5.8$ & $7.2\pm5.5$&... & 20.75\\
And XXII & 01:27:40.0 &+28:05:25 & $-1.8$ & $-6.5 \pm 9.9$ & 24.50 &  $-126.8\pm3.1$ & $3.54^{+4.16}_{-2.49}$&... & 20.56\\
NGC 147 & 00:33:21.1 & +48:30:32 & $-1.1 \pm 0.1$ & $-14.6 \pm 0.1$ & $24.15 \pm 0.09$ & $-193.1 \pm 0.8$\tablenotemark{e} & $16 \pm 1$\tablenotemark{e} & $0.356\pm 0.10$ & 20.35\\
NGC 185 & 00:38:58.0 & +48:20:15 & $-1.3 \pm 0.1$ & $-14.8 \pm 0.1$ & $23.95 \pm 0.09$ & $-208.8 \pm 1.1$\tablenotemark{e} & $24 \pm 1$\tablenotemark{e} & $0.120\pm 0.09$ & 20.12\\
NGC 205 & 00:40:22.1 & +41:41:07 & $-0.8 \pm 0.2$ & $-16.5 \pm 0.1$ & $24.58 \pm 0.07$ & $-246 \pm 1$\tablenotemark{f} & $35 \pm 5$\tablenotemark{f} & ...& 20.73\\
M32 & 00:42:41.8 & +40:41:55 & $-0.25$ & $-16.4 \pm 0.2$ & $24.53 \pm 0.21$ & $-196.9^{+ 5}_{-4.9}$\tablenotemark{g} & $29.9^{+ 5.2}_{- 4.6}$\tablenotemark{g} & $0.425\pm 0.32$ & 21.15 \\

\hline
\multicolumn{9}{p{15cm}}{a - \cite{McConnachie2012}, b - \cite{Tollerud2012}, unless otherwise noted, c - \cite{Ho2012}, d - \cite{Collins2013}, e - \cite{Geha2010}, f - \cite{Geha2006}, g- \cite{Howley2013}, h - \cite{Vargas2014}, i- Calculated in this work.}
\enddata
\label{tab:satellites}
\end{deluxetable*}

\subsection{Spectroscopic and Photometric Observations}
Our spectroscopic and photometric data were obtained over $\sim10$ years as part of the SPLASH survey. In this paper, we focus on the subset of SPLASH data that excludes the bright disk of M31. This includes fields targeting the dwarf spheroidals (dSphs), dwarf ellipticals (dEs), the smooth virialized halo, halo substructure, and M32. The extent of these observations are shown in Figure~\ref{fig:survey_map}. These data include 14143 stellar spectra taken with the DEIMOS multi-object spectrograph on the Keck-II 10m telescope. These spectra are spread across 151 individual DEIMOS masks, targeting $\sim60$ separate fields (red rectangles on Figure~\ref{fig:survey_map}). 

The SPLASH spectroscopic selection functions vary significantly from field to field, as the observations were conducted with specific science goals in mind that varied from field to field rather than following the strict guidelines of an overarching homogenous survey. As a result, we will not go into details regarding those selection functions in this section. Instead, we will discuss the selection functions to the extent that they affect the identification of carbon stars in Section \ref{selection functions}.

\subsubsection{dSphs}\label{dsphs}
SPLASH has observed 16 of the dSphs in the M31 system: And~I, And~II, And~III, And~V, And~VII, And~IX, And~X, And~XI, And~XII, And~XIII, And~XIV, And~XV, And~XVI, And~XVIII, And~XXI, and And~XXII. The properties of these dSphs relevant to this analysis are shown in Table~\ref{tab:satellites}.

The majority of these spectra were targeted using Washington photometry ($M$, $T_2$, and DDO51) taken with the Mosaic Camera on the Kitt Peak National Observatory (KPNO) 4m Mayall telescope. The DDO51 filter is centered on Mg absorption features (Mgb) that are highly dependent on surface gravity, and so allows for the discrimination of M31 giant stars from MW foreground dwarf stars \citep{Majewski2000, Gilbert2006}. We selected spectroscopic targets using the $M - {\rm DDO51}$ versus $M - T_2$ color-color diagram. The spectra in And~X were targeted using Sloan Digital Sky Survey (SDSS) imaging \citep{Adelman-McCarthy2006}. SDSS imaging was also used to supplement the Washington photometry imaging for And~II. The spectra in And~XV and And~XVI were targeted using archival Canada-France-Hawaii Telescope (CFHT) imaging. Finally, the spectra in And~XVIII and And~XXII were targeted using $B$- and $V$-band imaging from the Large Binocular Telescope (LBT).

All spectra were observed with the 1200 line mm$^{-1}$ grating with a central wavelength of $7800$~\AA. This configuration has a dispersion of $0.33~$\AA pixel$^{-1}$, and a wavelength range from H$\alpha$ to the Calcium {\sc II} triplet at 8500$~$\AA. The typical integration time was 3600 s per mask. 

For further descriptions of the original observations, we refer the reader to \citet[][And~XIV]{Majewski2007}, \citet[][And X]{Kalirai2009}, \citet[][And~I and III]{Kalirai2010}, \citet[][And II]{Ho2012}, and \citet[][the remainder]{Tollerud2012}. There are 3778 stellar spectra from dSph fields. All have associated Washington $M$, $T_2$ and DDO51 photometry (either used to target or taken after spectroscopic observations).

\subsubsection{dEs}
In addition to the dSphs, our dataset contains SPLASH observations of the three dEs of M31: NGC~147, NGC~185 and NGC~205. The properties of these dEs relevant to this analysis are shown in Table~\ref{tab:satellites}.

Spectra were targeted using CFHT CFH12K mosaic $R$ and $I$ band imaging \citep{Battinelli2004, Battinelli2004a}. Priority was assigned based on apparent $I$-band magnitude, with highest priority assigned to stars between $20.5 \le I_0 \le 21$. To minimize contamination by foreground MW dwarfs, stars were required to have $(R-I) > 0.2$. 

For the purposes of this work, we used the 12 masks designed to be observed in a conventional mode, with grating and exposure times matching the observations of other SPLASH fields (see \S~\ref{dsphs}). The final mask (n205-4m) was designed for use with a blocking filter centered on the CaT region at $8500$\AA. Since it did not cover the wavelength range necessary to identify carbon stars, it was excluded.

For further descriptions of the original observations, we refer the reader to \citet[][NGC~205]{Geha2006} and \citet[][NGC~147 and 185]{Geha2010}. In total, there are 1924 stellar spectra from dE fields with corresponding $R$ and $I$ magnitudes. In addition, these stars have CN and TiO narrow-band photometry, which will be discussed in greater detail later.

\subsubsection{M32}
SPLASH has also observed the compact elliptical (cE) galaxy M32. It's relevant properties are also listed in Table~\ref{tab:satellites}. The photometry for identifying spectroscopic targets was archival CFHT data imaged with MegaCam in the $g', r'$ and $i'$ bands. Heavy crowding means that the $g'$- and $r'$-band images were unreliable, so $i'$-band data drove the target selection. The greatest weight was given to unblended photometric sources with $20.5 \le I_o < 21$. 

All masks were observed with the same configuration described in \S~\ref{dsphs}. For further details on the observations, we refer the reader to \citet{Howley2013}. In total, there are 1418 stellar spectra with $i'$-band magnitudes.

\subsubsection{M31 Halo}
The remaining SPLASH fields target the halo of M31, including areas of known substructure \citep{Guhathakurta2006, Kalirai2006, Gilbert2007, Gilbert2009} and areas that are relatively smooth. The majority of these fields were targeted using KPNO Washington photometry, allowing for efficient separation of M31 giants and MW dwarfs \citep{Beaton2014PhD}. Additional photometry includes $V$- and $I$-band images taken with the William Herschel Telescope \citep{Zucker2007}, $V$- and $I$-band images taken with the Subaru Telescope's SuprimeCam \citep{Tanaka2010}, and $g'$ and $i'$ images (transformed to Johnson-Cousins $V$ and $I$) taken with CFHT's MegaCam. 

Spectra were observed using the standard SPLASH configuration (see \S~\ref{dsphs}). For further description of these observations, see \citet{Gilbert2012}, and references therein. There are 7023 stellar spectra in our halo fields; 3451 in fields containing substructure and 3572 in fields probing the smooth, virialized halo.

\subsection{Synthetic Photometry and Photometric Transformations}\label{synth phot}
The fields outlined above have been observed in a variety of filters: $M, T_2, V, I, R$, CN and TiO. To homogenize our dataset, we perform a series of photometric transformations to ensure that whenever possible stars have the equivalent of $V, I$ and $R$-band photometry. The Washington photometry filters $M$ and $T_2$ are nearly identical to Johnson Cousins $V$ and $I$. To convert from one to the other, we apply the transformation established by \citet{Majewski2000}. To transform $V$ to $R$, we use the transformation between $R-I$ and $V-I$ established for red and ``very red" stars by \citet{Battinelli2005}. This relationship is color dependent, with the break point at $V-I = 1.7$. After transforming $V-I$ to $R-I$, we add the $I$-band magnitude to extract $R$ on its own. To transform $R$ to $V$, we apply the same equations in the opposite direction. We do not further transform $V$ and $I$ into $M$ and $T_2$, simply because the Washington photometry filters are less commonly used in the literature and we do not require them for comparison. Typical uncertainties of the transformed photometry are 0.09~mag in $V$ and 0.02~mag in $R$. In contrast, the typical uncertainties of the raw photometry are 0.04~mag in $V$ and 0.05~mag in $R$.

We also calculate synthetic CN and TiO magnitudes. CN and TiO are narrow-band filters centered at $8120.5$\AA and $7778.4$\AA, respectively. The CN filter is centered on the CN band in carbon-stars and a continuum region of M-stars, while the TiO filter is centered on the TiO band in M-stars and a continuum region of C-stars. CN--TiO color can thus distinguish between C- and O-rich TP-AGB stars. Indeed many of the photometric surveys of AGB stars in the Local Group have used a broad band color and CN--TiO color to identify carbon stars \citep[the four-band photometry system, FBPS][]{Wing1971}. Broad-band photometry has too low spectral resolution to contain the detailed spectral information provided by narrow-band filters, so we instead compute synthetic CN and TiO magnitudes by weighting the spectra with CFHT/CH12k CN and TiO throughput curves \citep[see a full discussion of this method in][]{Hamren2015}. 

\subsection{Membership}\label{membership}

We would like to confirm whether the carbon stars identified in this paper are members of an M31 satellite, the M31 halo/extended disk, or the halo of the MW. To do this, we take advantage of the membership criteria established by the SPLASH team.

Membership in the halo and substructure fields is determined via the likelihood estimates from \citet{Gilbert2006}, which uses radial velocity, the equivalent width of the Na {\sc I} line at 8190.5\AA~(EW$_{\rm Na \sc I}$), position in $M$-DDO51 versus $M-T_2$ color-color space, position in $I$ versus $V-I$ color-magnitude space, and estimated [Fe/H] to determine the probably that a given star belongs to the M31 RGB or the MW foreground. Likelihood classes go from -3 (secure MW classification) through -1 (marginal MW classification) and 1 (marginal M31 classification) to 3 (secure M31 classification). In this work we will take those stars with likelihood greater than or equal to zero (i.e. those stars more likely to be associated with the M31 than the MW) to be M31 members. Likelihoods have also been calculated for many stars in the dSphs.

Membership in the M31 dSphs has been calculated by \citet[][And II]{Ho2012} and \citet[][the remainder]{Tollerud2012}. In And~II, the authors use a kinematical restriction, requiring member stars to be within $3\sigma$ of the systemic velocity of the dSph ($-228$ km s$^{-1} < v < -157$ km s$^{-1}$). They then apply a photometric and spectroscopic cut, $V-I < 2.5$ and EW$_{\rm Na I} < 4$, to further eliminate foreground MW contamination. In the remaining dSphs, the authors calculate membership probability using the distance of the star from the center of the dSph, the distance from the fiducial isochrones in $T_2$ versus $M - T_2$ space, the equivalent width of Na I, and the half-light radius of the dSph. 

In NGC~147 and NGC~185, membership criteria have been established by \citet{Geha2010}, using a modification of the method set forth in \citet{Gilbert2006}. The authors' metric uses line of sight velocity, EW$_{\rm Na\sc I}$, position with respect to isochrones in $I$ versus $V-I$ space, and  Ca {\sc II} triplet-based spectroscopic metallicity. 

Membership to NGC~205 is more difficult to determine: it's proximity to M31 leads to contamination from the M31 disk as well as the M31 and MW halos. As a rule of thumb, stars with velocities more negative than the systemic velocity of the dE ($v_{\rm sys} = -246 \pm 35$ km s$^{-1}$ ) are likely M31 halo stars. Stars with velocities much less negative than the systemic velocity of the dE are likely to be foreground contamination \citep{Geha2006}. While it is not used in \citet{Geha2006}, EW$_{\rm Na \sc I}$ can also be used to distinguish member stars from foreground dwarfs \citep[e.g.,][]{Gilbert2006}. Here we will apply the condition that EW$_{\rm Na \sc I} < 3$.

It is even more difficult to conclusively determine membership in M32, which is superimposed on the M31 disk. \citet{Howley2013} define M32 candidate members to be stars with $-275 \le v \le -125$ km s$^{-1}$. Stars with velocities more negative than this range likely belong to either M31's disk or inner spheroid. Stars with velocities less negative than this range are likely MW foreground.

\begin{figure*}
\centering
\includegraphics[width=6in]{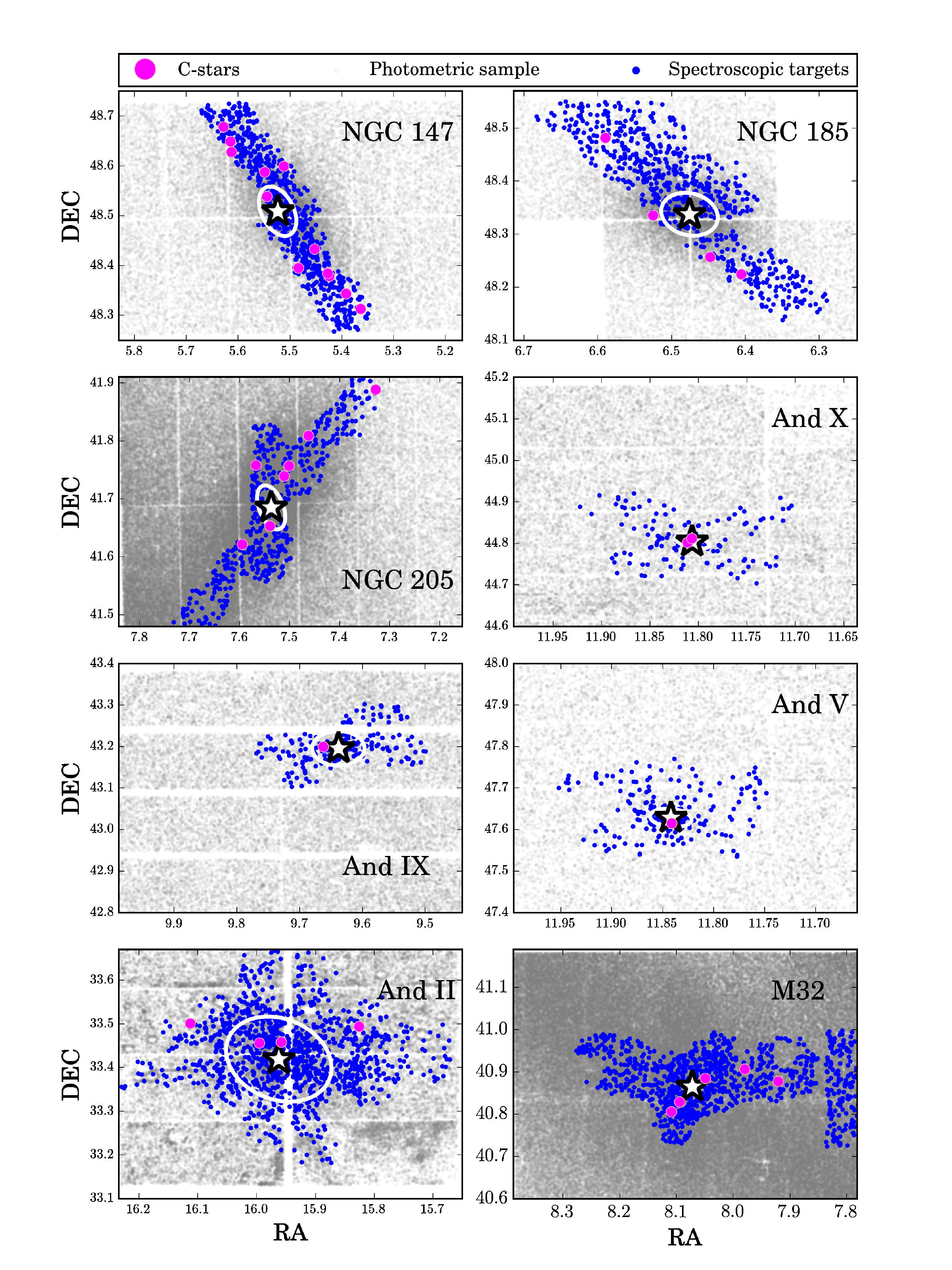}
\caption[Positions of carbon stars identified in M31 satellites]{Maps of carbon stars within their host satellite. Each panel shows the RA and DEC (in degrees) of the full photometric sample (plotted in grey) and the spectroscopic sample (overlaid in blue). Carbon stars (\S~\ref{cstars}) are plotted in pink. The fields have been stretched slightly to ensure that the full galaxy is visible in all panels. Each panel is annotated with the galaxy name, the galaxy center (white star), and an ellipse marking the half-light radius \citep[obtained from][]{McConnachie2012}. In several fields the half-light radius is so small as to be hidden behind the star marking the galaxy center.}
\label{fig:agb_map}
\end{figure*}

\section{The Carbon Star Sample}\label{cstars}

\subsection{Identification}

\begin{deluxetable*}{lcccclllcrc}
\tablewidth{0pt}
\tablecaption{Carbon stars in the satellites and halo of M31}
\tablehead{ ID & Field & Field Type & RA & DEC & $V$ & $I$ & $R$ & CN$-$TiO$^\ddagger$ & $v_{\rm helio}$ & Membership} \\ 
\startdata 

1000019 & And X & dSph & 01:06:35.21 & +44:48:06.3 & 22.31 & 21.84$^\dagger$ & 22.07$^\ddagger$ &  0.51 & $-184.4$ & And~X  \\ 
1000009 & And X & dSph & 01:06:33.77 & +44:48:43.8 & 21.59 & 20.87$^\dagger$ &21.22$^\ddagger$ &  0.59& $-175.1$& And~X \\ 
2002324 & And II & dSph & 01:17:13.14 & +33:30:04.6 & 21.52 & 19.72 &20.63$^\ddagger$ &  0.37& $-188.4$ & And~II\\ 
6004761 & And II & dSph & 01:15:50.50 & +33:29:37.9 & 22.49 & 21.19$^\dagger$ &21.83$^\ddagger$ &  0.54& $-215.8$ & And~II  \\ 
131 & And II & dSph & 01:16:28.58 & +33:27:28.6 &   $\ldots$&   $\ldots$ &  $\ldots$ &  0.61& $-192.9$ & And~II \\ 
2103321 & And II & dSph & 01:16:39.27 & +33:27:22.8 & 21.97 & 20.84$^\dagger$ &21.39$^\ddagger$ &  0.47& $-191.9$ & And~II  \\ 
6004761 & And II & dSph & 01:15:50.50 & +33:29:37.9 & 22.49 & 21.19$^\dagger$ &21.83$^\ddagger$ &  0.43& $-215.8$ & And~II \\ 
3005915 & And V & dSph & 01:10:16.79 & +47:36:56.5 & 21.91 & 20.52 &21.20$^\ddagger$ &  0.55& $-412.5$  & And~V \\ 
60001744 & And IX & dSph & 00:53:00.95 & +43:12:00.1 & 22.75 & 21.49$^\dagger$ &22.11$^\ddagger$ &  0.45& $-221.0$ & And~IX \\ 
240022 & M32 & cE& 00:42:34.43 & +40:53:04.7 &   $\ldots$ & 20.09 &  $\ldots$ &  0.39 & $-226.0$ & M32  \\ 
160934 & M32 & cE& 00:42:53.46 & +40:48:20.2 &   $\ldots$ & 20.46$\tablenotemark{*}$ &  $\ldots$ &  0.95&  $-2.6$ & MW/M31  \\ 
183365 & M32 & cE & 00:42:48.86 & +40:49:45.2 &   $\ldots$ & 20.02$\tablenotemark{*}$ &  $\ldots$ &  0.90& $-15.5$ & MW/M31\\
263107  &M32 & cE & 00:42:12.49 & +40:54:23.2 &   $\ldots$ & 20.48$\tablenotemark{*}$ &  $\ldots$ &  0.27& $-448.8$ & M31  \\ 
233791 & M32 & cE & 00:41:53.86 & +40:52:42.7 &   $\ldots$ & 20.55$\tablenotemark{*}$ &  $\ldots$ &  0.61& $-527.4$ & M31 \\ 
7003 & NGC 147 & dE & 00:32:45.10 & +48:22:44.4 & 23.65$^\ddagger$ & 20.90$^\dagger$ &22.42 &  0.71 & $-228.5$ & NGC~147  \\ 
12726 & NGC 147 & dE & 00:33:06.60 & +48:23:43.7 & 22.66$^\ddagger$ & 20.55$^\dagger$ &21.69 &  0.68 & $-235.2$& NGC~147\\ 
4380 & NGC 147 & dE & 00:32:23.20 & +48:18:46.5 & 22.26$^\ddagger$ & 20.03 &21.24 &  1.05 & $-255.4$ & NGC~147 \\ 
5306 & NGC 147 & dE & 00:32:33.40 & +48:20:36.7 & 22.84$^\ddagger$ & 20.61$^\dagger$ &21.83 &  0.74 & $-259.7$ & NGC~147 \\ 
7187 & NGC 147 & dE & 00:32:46.10 & +48:23:04.6 & 22.40$^\ddagger$ & 20.51$^\dagger$ &21.49 &  0.57 & $-269.4$ & NGC~147 \\ 
9245 & NGC 147 & dE & 00:32:55.30 & +48:25:59.5 & 22.82$^\ddagger$ & 21.32$^\dagger$ &22.05 &  0.43 & $-228.9$ & NGC~147 \\ 
20866 & NGC 147 &dE  & 00:33:30.20 & +48:35:15.5 & 22.31$^\ddagger$ & 20.13 &21.31 &  0.73 & $-218.7$ & NGC~147 \\ 
25366 & NGC 147 & dE & 00:33:53.40 & +48:37:41.7 & 23.05$^\ddagger$ & 21.50$^\dagger$ &22.26 &  0.51 & $-222.9$ & NGC~147 \\ 
25456 & NGC 147 & dE & 00:33:54.10 & +48:38:57.9 & 22.56$^\ddagger$ & 20.10 &21.46 &  0.47 & $-219.2$ & NGC~147 \\ 
25862 & NGC 147 & dE & 00:33:59.00 & +48:40:43.7 & 24.09$^\ddagger$ & 21.04$^\dagger$ &22.71 &  0.59 & $-198.1$ & NGC~147 \\ 
16315 & NGC 147 & dE & 00:33:16.80 & +48:35:57.9 & 22.57$^\ddagger$ & 20.65$^\dagger$ &21.66 &  0.53 & $-185.8$ & NGC~147 \\ 
os2 & NGC 147 & dE & 00:33:28.60 & +48:32:19.8 & 21.86$^\ddagger$ & 20.68$^\dagger$ &21.26 &  0.81& $-177.5$& NGC~147 \\
21924 & NGC 185 & dE & 00:39:39.26 & +48:28:55.2 & 23.10$^\ddagger$ & 21.49$^\dagger$ &22.28 &  0.40 & $-245.2$  & NGC~185 \\ 
3523 & NGC 185 & dE & 00:38:32.78 & +48:13:28.4 & 22.72$^\ddagger$ & 21.24$^\dagger$ &21.97 &  0.57 & $-233.6$ & NGC~185 \\ 
7625 & NGC 185 & dE & 00:38:47.89 & +48:15:25.7 & 21.82$^\ddagger$ & 20.04 &20.93 &  0.52 & $-254.7$& NGC~185 \\ 
17901 & NGC 185 & dE & 00:39:15.88 & +48:20:08.2 & 21.86$^\ddagger$ & 20.03 &20.97 &  0.40 & $-220.5$ & NGC~185 \\ 
22282 & NGC 205 & dE & 00:40:32.10 & +41:45:29.4 & 22.69$^\ddagger$ & 20.51 &21.69 &  0.63 & $-243.3$ & NGC~205 \\ 
23520 & NGC 205 & dE & 00:40:10.67 & +41:45:24.7 & 22.91$^\ddagger$ & 20.59 &21.86 &  0.71 & $-228.3$ & NGC~205 \\ 
2186&NGC 205 & dE & 00:39:15.05 & +41:53:19.8 & 22.62$^\ddagger$ & 20.66 &21.69 &  0.74 & $-195.9$& NGC~205 \\
5100 & NGC 205 & dE & 00:39:58.28 & +41:48:33.2 & 22.99$^\ddagger$ & 20.55 &21.89 &  0.63 & $-162.9$ & NGC~205 \\ 
26251 & NGC 205 & dE & 00:40:13.88 & +41:44:21.4 & 22.72$^\ddagger$ & 20.86$^\dagger$ &21.82 &  0.74 & $-183.9$ & NGC~205 \\ 
80542 & NGC 205 & dE & 00:40:23.04 & +41:39:12.8 & 22.17$^\ddagger$ & 20.04 &21.19 &  0.44 & $-276.3$ & NGC~205 \\ 
83287 & NGC 205 & dE & 00:40:40.70 & +41:37:17.3 & 22.45$^\ddagger$ & 20.54 &21.53 &  0.73 & $-423.8$ & M31 \\ 
507513 & Northeast shelf & substruct & 00:50:15.66 & +41:42:07.7 & 23.08 & 21.44 &  $\ldots$ &  0.68 & $-218.0$ & M31 \\ 
137459 &Northeast shelf & substruct & 00:52:14.21 & +42:08:55.5 & 23.03 & 21.29 &  $\ldots$ &  0.60 & $-315.2$  & M31 \\ 
170544 & Northeast shelf & substruct & 00:52:45.74 & +42:15:24.8 & 22.91 & 21.01 &  $\ldots$ &  0.89 & $-376.1$ & M31 \\ 
17901 & Northwest shelf & substruct & 00:34:42.85 & +42:25:21.3 & 22.35 & 20.81 &21.56 &  0.50 & $-365.4$ & M31 \\
\hline
\multicolumn{10}{p{10cm}}{ $^{\rm *}$~$i'$ photometry calibrated to match Johnson-Cousins $I$ } \\
\multicolumn{10}{p{10cm}}{$^\dagger$~Fainter than $I_{\rm TRGB}$} \\
\multicolumn{10}{p{10cm}}{$^\ddagger$~Synthetic or transformed photometry, as per \S~\ref{synth phot}} 
\enddata 
\label{tab:cstars}
\end{deluxetable*}

The optical spectra of carbon stars are distinguished by prominent CN features at $\sim7000-8200$\AA, and C$_2$ bands at $\sim6100-6600$\AA. Previous authors have used the CN band to identify carbon stars, either by using narrow-band filters centered on CN and its oxygen-rich counterpart TiO \citep[e.g.,][]{Nowotny2003, Battinelli2004, Battinelli2004a, Wing2007}, or by cross-correlating spectra against templates with and without CN \citep{Hamren2015}.

In this work, we identify carbon stars using the spectroscopic classification statistic demonstrated by \citet{Hamren2015} in the disk of M31. This method identifies carbon stars by cross correlating the spectrum in question with a suite of Milky Way templates. Spectra that are best fit by a carbon star template are flagged as likely carbon stars, and visually examined for final confirmation. We refer the reader to \citet{Hamren2015} for further details about the template observations and testing of the classification metric. 

To make use of the high S/N carbon templates, which were observed with DEIMOS's 600 line grating, we first rebin our spectra to match the template spectra's 0.65\AA~pixel$^{-1}$ dispersion. We then apply the classification statistic to the full SPLASH satellite/halo sample of 14143 stars and identify 41 carbon stars. The full list of carbon stars, including their SPLASH ID number, their position, magnitude, and velocity, is presented in Table~\ref{tab:cstars}. For the carbon stars found in the satellites, their position within the galaxy (whose center is denoted by a white X) is shown in Figure~\ref{fig:agb_map}. These maps are omitted for stars in the halo fields, as there is no discernible structure or center to use as a useful reference point. 

Using the membership criteria outlined in \S~\ref{membership}, we find that all carbon stars in the dSph fields are unambiguous members of their respective satellite. The same is true for the carbon stars in NGC~185. Only two of the 12 carbon stars identified in NGC~147 fields satisfy the full set of membership criteria outlined in \S~\ref{membership}. However, if we take into account the rotational velocity of the dE and radial-dependent velocity dispersion \citep{Geha2006}, then all the stars are within $3\sigma$ of the systemic velocity. We will thus mark all carbon stars in NGC~147 as members.

Four of the five carbon stars identified in NGC~205 have kinematics consistent with the dE. Star 83287 has a velocity of $-423.8$ km s$^{-1}$, making it more likely to be a member of the M31 halo. In M32, only star 240022 has a velocity consistent with the cE. Stars 160934 and 183365 have velocities that may indicate either foreground carbon dwarfs or stars in the tail of the M31 spheroid's kinematic distribution. Given that these masks are located on a relatively high surface brightness part of M31, picking up a foreground star (let alone a rare dC star) is unlikely. These two stars are thus more likely to belong to the M31 spheroid. Stars 263107 and 233791 have velocities consistent with either the M31 spheroid or disk.

We also identify four carbon stars in fields with known substructure, associated with M31's Northeast or Western Shelves. These shelves are debris from the Giant Southern Stream, are are known to contain AGB stars \citep[e.g.][]{Tanaka2010}.

\subsection{Selection Functions}\label{selection functions}

The presence, or lack thereof, of carbon stars in a particular field is highly dependent on the spectroscopic selection function. The identification of carbon stars was never one of the science goals of the SPLASH survey, and so the regions in color-magnitude space in which they are typically found were not always prioritized. 

To estimate the number of carbon stars we might expect to observe in each field, we begin with the assumption that our carbon stars are all TP-AGB stars brighter than the tip of the red giant branch (TRGB). To estimate the total number of TP-AGB stars in each field (N$_{AGB}$) we count the number of spectral targets brighter than the TRGB that have $V-I > 1.5$ (equivalently, $R-I > 0.73$). These limits are based on the region in color-magnitude space where carbon-rich TP-AGB stars are typically found, with a slightly bluer color limit designed to encompass the majority of our sample. 

In the satellites, we can calculate the $I$-band magnitude of the TRGB using the calibration from \citet{Bellazzini2004} adjusted by the distance moduli listed in Table~\ref{tab:satellites}. This calibration requires [M/H] measurements, so we convert the [Fe/H] measurements in Table~\ref{tab:satellites} using Equation~1 from \citet{Ferraro1999}. We use the [$\alpha$/Fe] measurements from \citet{Vargas2014} where available, and assume [$\alpha$/Fe]$ = 0.28$ \citep[the same value used by][]{Ferraro1999} for the remainder. Finally, we correct the calculated $I_{\rm TRGB}$ for foreground extinction using the dust maps from \citet{Schlafly2011}. Our calculated $I_{\rm TRGB}$ are listed in Table~\ref{tab:satellites}. Because the $i'$ photometry of M32 was specifically calibrated such that $i'_{\rm TRGB}$ and $I_{\rm TRGB}$ are equivalent, our calculated $I_{\rm TRGB}$ for M32 is still comparable to the $i'$ photometry. For the halo fields, we assume that $I_{\rm TRGB}$ is located at $M_{\rm bol} = -4$, and adopt the distance modulus of M31 \citep[24.47,][]{Stanek1998}, giving us $I_{\rm TRGB} = 20.47$.

In the fields with potential TP-AGB stars, we can estimate how many carbon stars we would expect to observe using a theoretical C/M ratio --- the number ratio of carbon- to oxygen-rich TP-AGB stars. We calculate the C/M ratio in each field using the well-calibrated relationship between log(C/M) and [Fe/H] from \citet{Cioni2009}. In satellite fields, we adopt the [Fe/H] values listed in Table~\ref{tab:satellites}. In halo fields, we compute the projected line-of-sight distance from M31, and then derive [Fe/H] value using the metallicity gradient from \citet{Gilbert2014}. 

The probability of detecting a C-star in a TP-AGB population (assuming that the region in color and magnitude space described above is populated entirely by C- and M-type AGB) is 
\begin{equation}
P_c = {1\over{(C/M)^{-1} + 1}}
\end{equation}
Multiplying this probability by the number of likely TP-AGB stars gives us the number of carbon stars expected in each field. 

Table~\ref{tab:selection} outlines this analysis. For each field, it lists the theoretical C/M ratio, the number of TP-AGB stars (N$_{AGB}$), the number of predicted carbon stars (N$_{\rm CP}$), and the number of observed carbon stars (N$_{\rm CO}$). We also include the projected radial distance for all fields, although it was only used in the analysis of halo fields.

There are 25 fields in which there are zero predicted C-stars and zero observed C-stars. Of these 25, five are dSphs (And~XI, And~XIII, And~XV, And~XVI and And~XXII). These five dSphs are some of the least massive of our sample, where the majority of bright stars would have been put on a mask. There are four fields in which we do identify some carbon stars despite there being none predicted; And~X, NE2, NE4, and NE1. These carbon stars are either fainter than the TRGB or bluer than $V-I = 1.8$, and will be discussed in detail later in this paper.  

In most of the remaining fields, we observe roughly the number of carbon stars that we expect (to within a factor of two). However, in And~II, NGC~147, NGC~185 and NGC~205 we observe far more carbon stars than are predicted, despite the fact that our calculated C/M ratios match those in the literature. As in And~X and the Northeast Shelf fields, many of these stars are fainter than the TRGB and/or bluer than $V-I = 1.8$. In And~I and And~III, we predict several carbon stars ($\sim4.5$ in both fields), but do not observe any. 

\begin{longtable*}{lcccccc}
\tablewidth{0pt}
\tablecaption{Selection Function Data}
\tablehead{Field & $N_{\rm masks}$ & $D_{\rm M31, proj}^a$ & Theoretical C/M & N$_{\rm AGB}^b$ & N$_{\rm C,~pred}^c$ & N$_{\rm C,~obs}^d$ \\
			& & (kpc) & & & &}
 SE8 & 1 &   3.8 &  0.001 & 0 &  0.00 & 0 \\
 SE9 & 1 &   6.0 &  0.001 & 0 &  0.00 & 0 \\
 f109 & 1 &   8.9 &  0.002 & 1 &  0.00 & 0 \\
 f1 & 2 &  12.2 &  0.002 & 1 &  0.00 & 0 \\
 NW1V & 1 &  13.1 &  0.002 & 1 &  0.00 & 0 \\
 f116 & 1 &  13.2 &  0.002 & 1 &  0.00 & 0 \\
 f115 & 1 &  14.5 &  0.002 & 3 &  0.01 & 0 \\
 f207 & 1 &  16.3 &  0.002 & 0 &  0.00 & 0 \\
 NW1dV & 1 &  18.0 &  0.003 & 2 &  0.01 & 0 \\
 25kpc & 1 &  20.0 &  0.003 & 0 &  0.00 & 0 \\
 NE3 & 1 &  20.6 &  0.003 & 0 &  0.00 & 0 \\
 f123 & 1 &  21.0 &  0.003 & 0 &  0.00 & 0 \\
 f2 & 2 &  21.3 &  0.003 & 2 &  0.01 & 0 \\
 NW2V & 2 &  21.5 &  0.003 & 1 &  0.00 & 0 \\
 NE1 & 1 &  22.8 &  0.003 & 0 &  0.00 & 1 \\
 NW2dV & 1 &  24.2 &  0.004 & 2 &  0.01 & 0 \\
 M32 & 6 &   5.7 &  0.004 & 0 &  0.00 & 5 \\
 NE6 & 1 &  24.8 &  0.004 & 0 &  0.00 & 0 \\
 f130 & 2 &  25.3 &  0.004 & 3 &  0.01 & 0 \\
 NW3V & 1 &  26.8 &  0.004 & 1 &  0.00 & 1 \\
 a0 & 3 &  29.8 &  0.005 & 4 &  0.02 & 0 \\
 NE4 & 1 &  32.9 &  0.006 & 0 &  0.00 & 1 \\
 g1 & 1 &  34.5 &  0.006 & 0 &  0.00 & 0 \\
 a3 & 3 &  35.4 &  0.006 & 4 &  0.03 & 0 \\
 mask4 & 1 &  36.8 &  0.007 & 0 &  0.00 & 0 \\
 f135 & 1 &  37.9 &  0.007 & 0 &  0.00 & 0 \\
 NE2 & 1 &  39.1 &  0.008 & 0 &  0.00 & 1 \\
 A240 & 3 &  55.2 &  0.017 & 0 &  0.00 & 0 \\
 A338 & 3 &  56.1 &  0.018 & 1 &  0.02 & 0 \\
 m4 & 5 &  56.6 &  0.018 & 4 &  0.07 & 0 \\
 a13 & 4 &  58.0 &  0.019 & 5 &  0.10 & 0 \\
 NW9V & 2 &  65.4 &  0.028 & 1 &  0.03 & 0 \\
 n205 & 3 &  32.6 &  0.056 & 34 &  1.79 & 7 \\
 a19 & 4 &  79.5 &  0.056 & 1 &  0.05 & 0 \\
 A220 & 3 &  85.2 &  0.075 & 2 &  0.14 & 0 \\
 m6 & 5 &  85.3 &  0.075 & 0 &  0.00 & 0 \\
 A310 & 3 &  87.1 &  0.082 & 1 &  0.08 & 0 \\
 A040 & 3 &  90.3 &  0.096 & 0 &  0.00 & 0 \\
 b15 & 5 &  91.9 &  0.104 & 1 &  0.09 & 0 \\
 NW13dV & 1 &  92.3 &  0.106 & 0 &  0.00 & 0 \\
 NW14V & 1 &  95.5 &  0.125 & 0 &  0.00 & 0 \\
 NW15V & 1 & 100.3 &  0.158 & 0 &  0.00 & 0 \\
 N147 & 4 & 101.4 &  0.242 & 7 &  1.36 & 12 \\
 m8 & 2 & 116.8 &  0.356 & 1 &  0.26 & 0 \\
 N185 & 5 &  97.4 &  0.643 & 4 &  1.57 & 4 \\
 d7 & 2 & 220.1 &  1.050 & 3 &  1.54 & 0 \\
 d1 & 2 &  44.8 &  1.342 & 8 &  4.58 & 0 \\
 d5 & 4 & 109.8 &  2.798 & 2 &  1.47 & 1 \\
 m11 & 4 & 158.6 &  2.827 & 2 &  1.48 & 0 \\
 A170 & 2 & 159.9 &  3.014 & 0 &  0.00 & 0 \\
 d2 & 12 & 140.5 &  3.403 & 11 &  8.50 & 5 \\
 A080 & 2 & 164.8 &  3.838 & 1 &  0.79 & 0 \\
 A305 & 2 & 169.5 &  4.842 & 0 &  0.00 & 0 \\
 d3 & 3 &  68.4 &  6.758 & 5 &  4.36 & 0 \\
 d18 & 1 & 113.0 &  7.453 & 1 &  0.88 & 0 \\
 d22 & 3 & 245.1 &  7.453 & 0 &  0.00 & 0 \\
 d15 & 2 &  93.7 &  7.453 & 0 &  0.00 & 0 \\
 d13 & 5 & 126.8 & 12.165 & 0 &  0.00 & 0 \\
 d10 & 2 &  76.7 & 14.091 & 0 &  0.00 & 2 \\
 d11 & 1 & 102.2 & 19.855 & 0 &  0.00 & 0 \\
 d16 & 2 & 138.7 & 32.407 & 0 &  0.00 & 0 \\
 d12 & 1 & 114.0 & 32.407 & 1 &  0.97 & 0 \\
 d9 & 2 &  36.8 & 52.894 & 5 &  4.91 & 1 \\
 d14 & 1 & 159.8 & 70.968 & 1 &  0.99 & 0 \\
 \hline
 \multicolumn{6}{p{8cm}}{$^a$~The projected distance (in kpc) of each field from the center of M31} \\
 \multicolumn{6}{p{8cm}}{$^b$~The number of likely AGB stars observed in each field (see text for details)} \\
 \multicolumn{6}{p{8cm}}{$^c$~The \textit{predicted} number of carbon stars in each field given the total number of observed AGB stars and the theoretical C/M ratio} \\
 \multicolumn{6}{p{8cm}}{$^d$~The \textit{observed} number of carbon stars in each field }
\label{tab:selection}

\end{longtable*}

\section{Photometric properties}\label{photometry}

In this section we will examine the photometric properties of our sample of 41 carbon stars. Several of these stars were identified in the previous section as being fainter than the TRGB. They are located in And~II, And~IX, And~X, and all three dEs, and are noted in Table~\ref{tab:cstars}. As these faint (sub-TRGB) carbon stars may be extrinsic in origin, we will distinguish between them and the bright (super-TRGB) sample.

\begin{figure}
\centering
\includegraphics[width=3in]{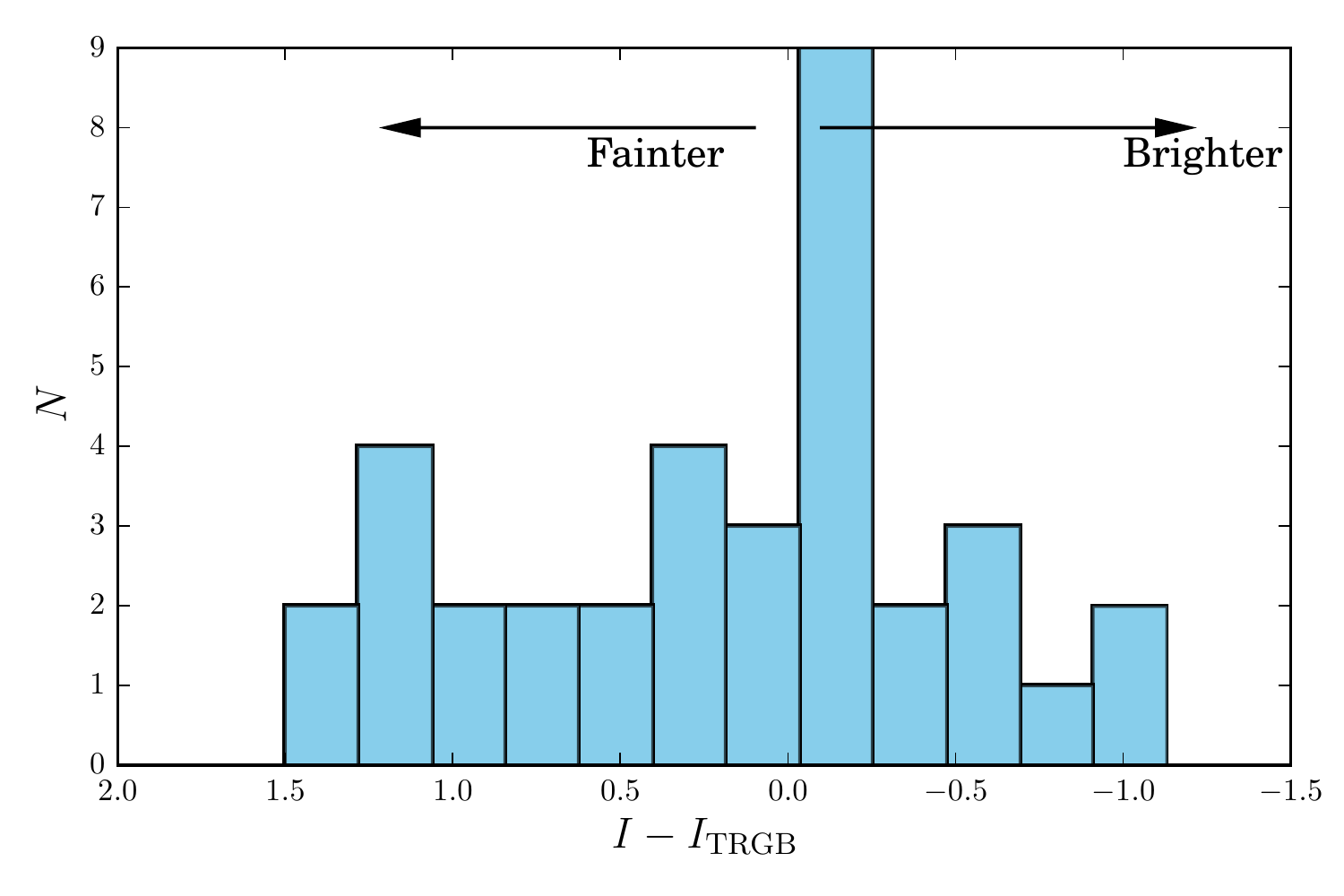}
\caption[Distribution of C-star magnitudes with respect to the TRGB]{Distribution of carbon star $I$-band magnitude in the dEs and dSphs of our sample with respect to the TRGB of their host galaxy; 42\% are brighter than the TRGB and 58\% are fainter.}
\label{fig:below_trgb}
\end{figure}

The distribution of C-star magnitudes with respect to their host satellite's TRGB is shown in Figure~\ref{fig:below_trgb}. We restrict this figure to those C-stars within dSphs and dEs, because their distance moduli and thus TRGB estimates are far more certain than in the halo fields. Only 13 ($42\%$) of the carbon stars in the M31 satellites are brighter than the TRGB. The uncertainty in this fraction is dominated by the uncertainties on the distance moduli, which can be as high as 0.2 magnitudes. Typical uncertainties on the $I$-band magnitudes of the carbon stars, on the other hand, is $0.02 \pm 0.01$ mag.

The bright carbon stars are concentrated close to the TRGB, with the brightest observed C-star less than one magnitude away. This may be a selection effect, as very bright stars were occasionally given low targeting priority to avoid possible foreground contamination. The remaining 18 carbon stars are fainter than the TRGB by between 0.13 and 1.5 magnitudes. Given the small number of carbon stars in each satellite, it is difficult to compare the observed carbon star luminosity function with the observed SPLASH luminosity function, and we cannot tell if carbon stars are represented down to the detection limit. 

\begin{figure}
\centering
\includegraphics[width=3in]{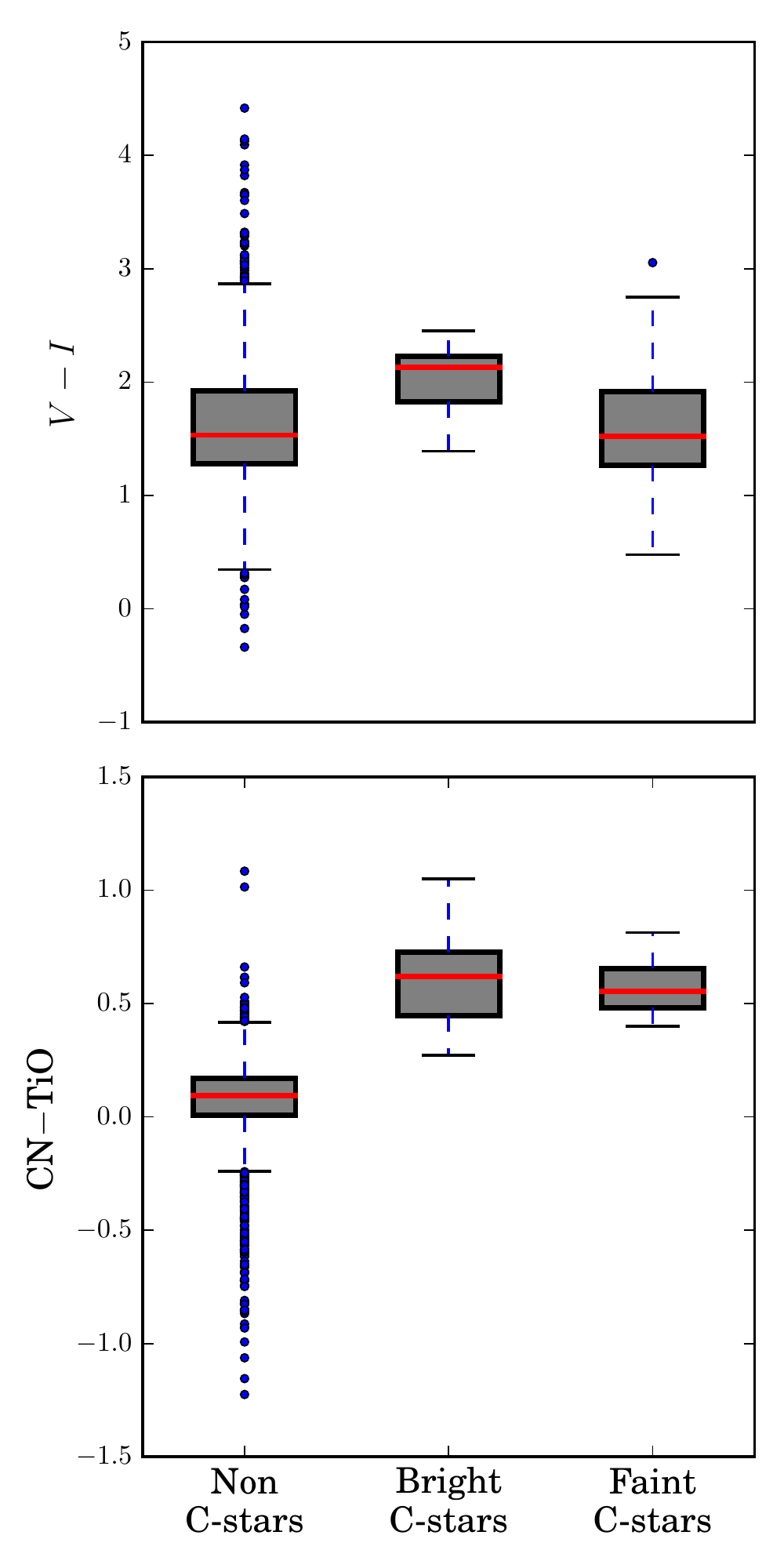}
\caption[Distribution of $V-I$ and CN$-$TiO]{Distributions of non-carbon stars (left), C-stars brighter than the TRGB (center), and C-stars fainter than the TRGB (right) in $V-I$ (top) and CN$-$TiO (bottom). Solid red lines indicate the median of the distribution, grey boxes encompass the quartiles, and the whiskers extend 1.5 times the interquartile range past the quartiles. Outliers are displayed as circles. The bright C-stars are redder in $V-I$ than the faint C-stars, despite having comparable CN--TiO colors (i.e. CN band strength).}
\label{fig:boxplot}
\end{figure}

Figure~\ref{fig:boxplot} shows the distribution of colors, rather than magnitudes: $V-I$ color, representing the slope of the continuum, and CN--TiO color, representing the strength of the CN features at $\sim7900$\AA. We compare stars with no carbon features, carbon stars brighter than the TRGB and carbon stars fainter than the TRGB. The bulk of the faint C-stars are bluer in $V-I$ than the bright C-stars, with a distribution more like that of the non-carbon stars. However the faint carbon stars have comparable CN--TiO colors to the bright C-stars, both of which are considerably higher than the non-carbon stars. This indicates comparable strength of carbon features. The distribution of CN--TiO color for non-carbon stars overlaps slightly with the distributions of CN--TiO for both groups of carbons tars. However visual examination of the spectra shows that this to due to artifacts introduced by the sky subtraction process, and not the presence of carbon stars we failed to identify.

\begin{figure*}[t]
\centering
\includegraphics[width=7in]{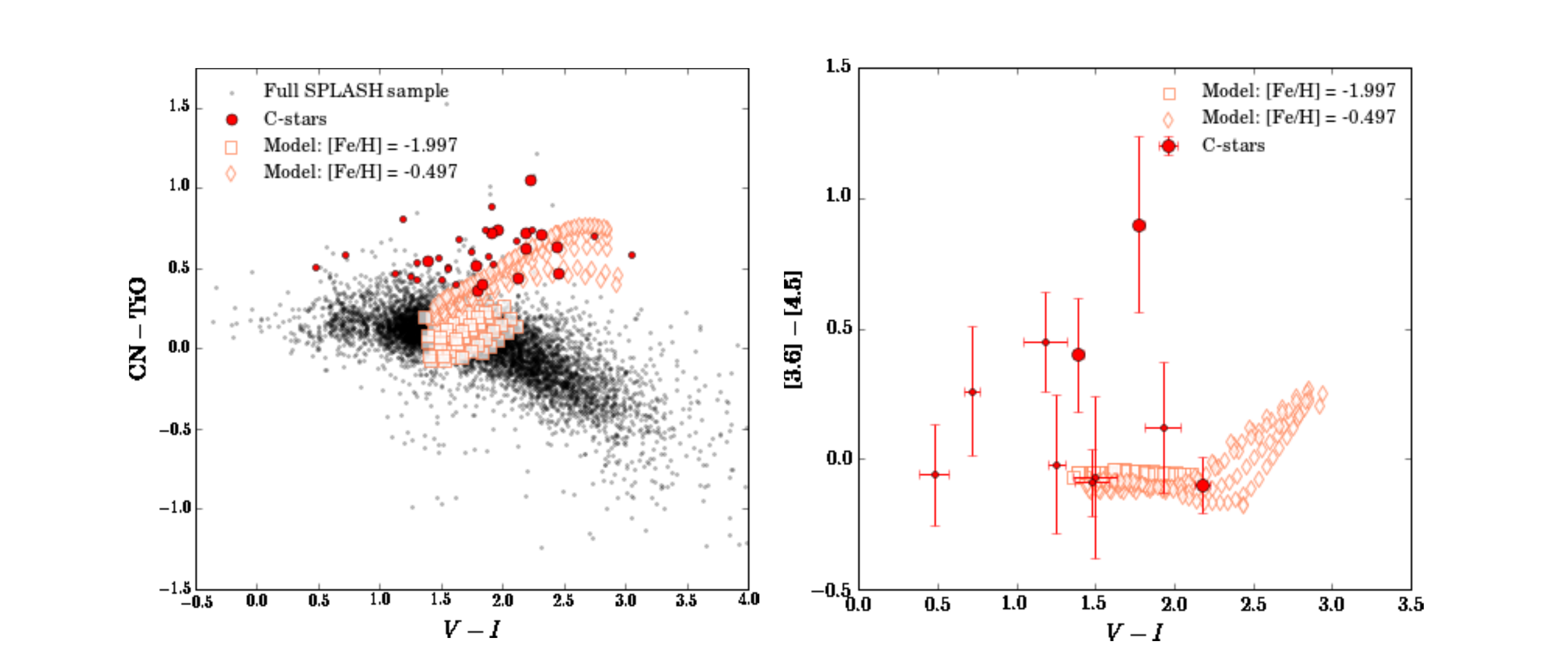}
\caption[Color-color diagrams]{ Color-color diagrams of observed and synthetic carbon stars. The left panel shows synthetic CN--TiO versus SPLASH $V-I$ color, and the right panel shows DUSTiNGS [3.6]--[4.5] versus SPLASH $V-I$. In both panels SPLASH carbon stars are shown as blue points sized by whether the star is fainter (small points) or brighter (large points) than $I_{\rm TRGB}$. Diamonds and squares represent the dust-free hydrostatic models from \cite{Aringer2016}. On the left panel we also the full SPLASH sample as small grey points. }
\label{fig:color_color}
\end{figure*} 

In addition to the one-dimensional distributions of color in Figure~\ref{fig:boxplot}, it is also constructive to look at these distributions in two dimensions. Figure~\ref{fig:color_color} shows two color-color diagrams. The left panel plots CN--TiO versus $V-I$, with the full SPLASH sample (black points) overlaid with the dust-free, hydrostatic synthetic photometry from \citet{Aringer2016}. The carbon stars are plotted as blue points, with faint and bright stars differentiated by point size. The carbon stars are fairly consistent with the models, which indicates that they are not particularly dusty. In general, the bright C-stars (larger points) are a better fit, as many of the fainter carbon stars (smaller points) are bluer than the models. However there is not a clear dichotomy; some of the faint C-stars fit the models better than some of the bright C-stars, and visa versa. There are no stars substantially redder than the models in $V-I$. This is likely a selection effect, as the photometry with which SPLASH spectroscopic targets were selected is less complete at these colors. Indeed the sample of non-carbon stars falls off dramatically at $V-I > 3$ as well.

The right panel of Figure~\ref{fig:color_color} takes advantage of the fact that many of our dSph and dE fields have also been imaged by the Survey of Dust in Nearby Galaxies with \textit{Spitzer} \citep[DUSTiNGS][]{Boyer2015}. DUSTiNGS imaged 50 nearby dwarf galaxies at 3.6 and 4.5 $\mu$m using the InfraRed Array Camera \citep[IRAC][]{Fazio2004} on the \textit{Spitzer Space Telescope}. In total there are 1749 stars with SPLASH spectra that fall within the DUSTiNGS footprints, 11 of which are carbon stars. The 3.6 and 4.5 $\mu$m photometry of these 11 stars is shown in Table~\ref{tab:mir}. 

\begin{deluxetable}{lcccc}
\tablewidth{0pt}
\tablecaption{Carbon stars in the DUSTiNGS sample}
\tablehead{ID & Mask & $3.6$ & $4.5$ }\\
\startdata
1000019 & d10\_1 & 18.87 & 18.93 \\
1000009 & d10\_2 & 19.64 & 19.38\\
131 & d2\_12 & 20.48 & 20.15\\
3005915  & d5\_1 & 19.18 & 18.78\\
60001744 & d9\_1 & 19.97 & 19.99\\
9245 & N147\_2 & 19.86 & 19.93\\
20866 & N147\_3 & 18.40 & 18.50\\
16315 & N147\_4 & 19.69 & 19.57\\
os2 & N147\_4 & 17.58 & 17.13\\
3523 & N185\_3  & 18.75 & 18.84\\
7625 & N185\_3 & 20.31 & 19.41
\enddata
\label{tab:mir}
\end{deluxetable}

The right panel of Figure~\ref{fig:color_color} displays [3.6]--[4.5] versus $V-I$. Included are the dust-free hydrostatic models from \citet{Aringer2016} and the photometry of the ten carbon stars with optical and MIR photometry. We have omitted the matched DUSTiNGS-SPLASH photometry for the full SPLASH sample, because it is highly biased. The SPLASH sample extends far below the DUSTiNGS detection limit, meaning that many of the matched stars are some of the faintest in the DUSTiNGS sample. Stars so close to the detection limit are more often observed when random noise makes them brighter and as the detection limit in 4.5$\mu$m is higher than in 3.6$\mu$m this translates to artificial reddening. Most TP-AGB C-stars should not be affected by this bias.

The fit between the observed C-star photometry and the synthetic photometry is worse in [3.6]--[4.5] versus $V-I$ color-color space than it is in CN--TiO versus $V-I$. As before, while the faint carbon stars (small points) are often a poor fit to the models, there are some faint C-stars that fit the models better than the bright C-stars, and visa versa. Some of the mismatch may be due to internal stellar dynamics, as the optical and MIR photometry was taken at very different times. A one dex shift in $V-I$, at least for the bright C-stars, would lead to a far better fit to the models. This shift is well within the variation amplitude of a carbon-rich variable star, and models predict larger color amplitudes in the optical than at longer wavelengths  \citep{Nowotny2011, Nowotny2013}. That said, none of these stars (faint C-stars included) appear in the DUSTiNGS variable star/extreme-AGB star catalog \citep{Boyer2015} (though with only two DUSTiNGS epochs this does not definitively rule out variability). Pulsation is unlikely to be enough to explain the position of the three bluest faint-carbon stars. That these are such a poor fit to the models suggests that they are not AGB stars at all.

\section{Spectroscopic Properties}\label{spectroscopy}

\begin{figure*}
\centering
\includegraphics[width=7in]{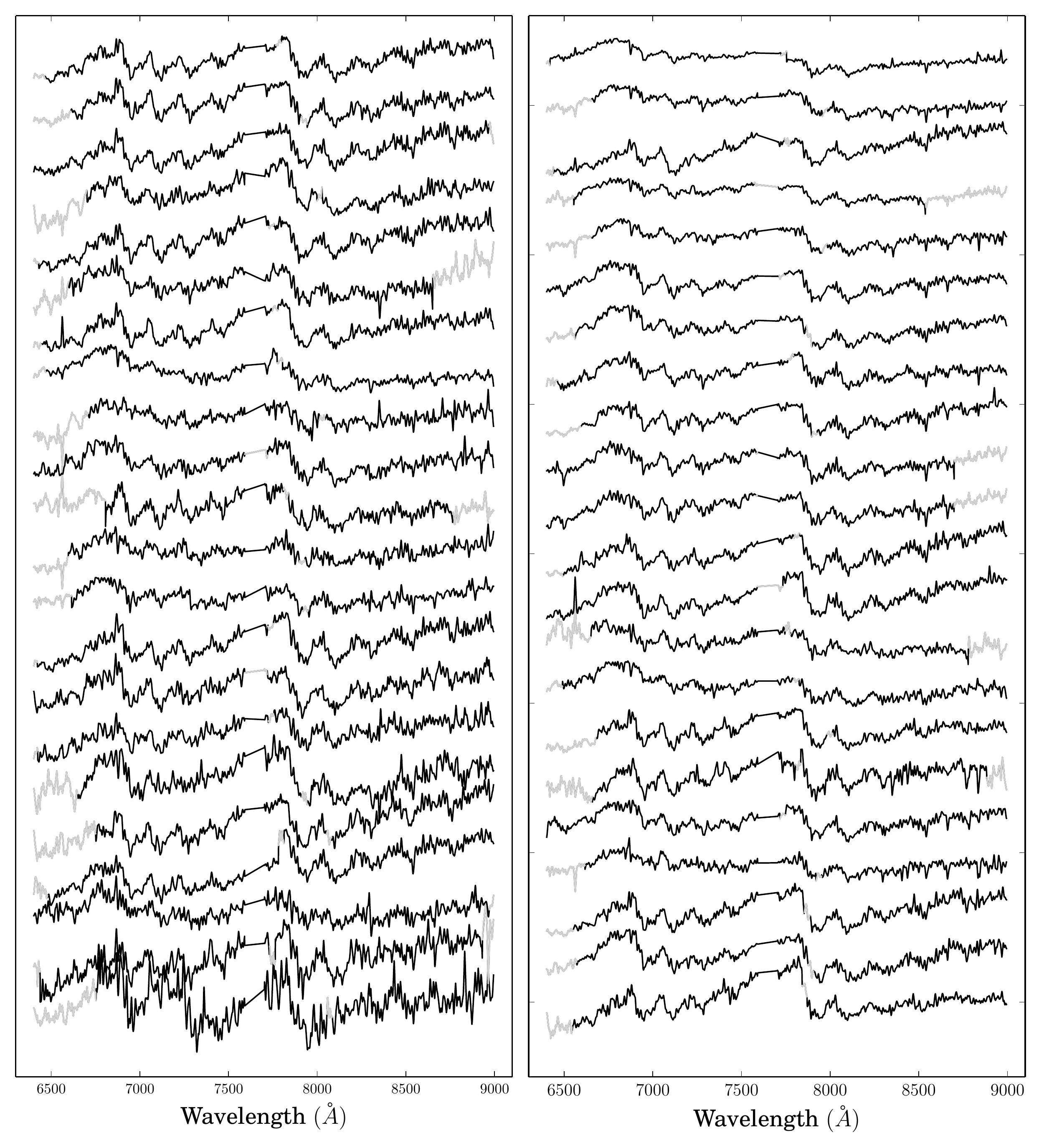}
\caption[PCA-reconstructed C-star spectra]{Spectra of all carbon stars in our sample, with regions of missing data reconstructed with PCA. Spectra have been smoothed by a Gaussian with $\sigma=1.95$\AA, and the telluric A-band has been replaced with a linear interpolation. Sections of the spectra that are reconstructed using eigenvectors are highlighted in grey, while the ``raw" spectra are plotted in black.}
\label{fig:all_cstars}
\end{figure*}

The optical spectra of carbon stars are characterized by strong bands of CN, C$_2$, and CH. Of these three molecules, only CN and C$_2$ are visible within the wavelength range covered by the SPLASH spectra (H$\alpha$ to the Ca {\sc II} triplet). These features do not vary as strongly with temperature as their counterparts in oxygen-rich stars (namely TiO), and, as a result, optical carbon star spectra are remarkably uniform \citep[e.g.,][]{VanLoon2005}. 

To identify the fundamental components of our spectra, we apply principal component analysis (PCA) to derive ``eigenspectra." Broadly, PCA is a technique for dimensionality reduction, which converts a set of observations of potentially correlated properties to an orthogonal, uncorrelated, basis set. It has been used to classify the spectra of galaxies \citep{Connolly1995}, quasars \citep{Yip2004}, and stars \citep{McGurk2010}, in addition to being widely used by other disciplines. 

Briefly, eigenspectra are computed in much the same way as standard eigenvectors. We begin with a symmetric correlation matrix C, such that C$_{ij}$ is the normalized scalar product of spectra $i$ and $j$. We then find the matrix U, such that  
\begin{equation}
U^TCU = \Lambda
\end{equation}
where C is the correlation matrix, and $\Lambda$ is a diagonal matrix of eigenvalues. The $i$-th column of U contains the $i$-th eigenspectrum. The ``weight" of each eigenspectrum in the observed spectra is called the eigencoefficient, and is obtained by projecting the observed spectra onto the eigenspectra. 

In the end, each observed spectrum can be described by the following expression
\begin{equation}
x_i = \mu + \sum\limits_{j=1}^{N}a_{ij}e_j
\end{equation}
where $x_i$ is the spectrum in question, $\mu$ is the mean spectrum, $e_j$ are the eigenspectra, and $a_{ij}$ are the eigencoefficients. 

To compute the eigenspectra, eigenvalues, and coefficients of our set of carbon stars, we use the \textit{iterative} PCA formalism set forth by \citet[][hereafter Y04]{Yip2004} in the \texttt{astroML} Python library \citep{astroML}. Iterative PCA allows us to use successively regenerated eigenvectors to fill holes of missing data, essential for our spectra, which have variable wavelength coverage. However, iterative PCA cannot reconstruct regions in which no spectra have data. This is an issue for the telluric A-band, which was improperly corrected during data processing. Because the telluric A-band contains no stellar information, we remove it ($7591-7703$\AA) and linearly interpolate across the gap. We then smooth the spectra with a Gaussian of $\sigma=1.95$\AA and perform iterative PCA with l2 normalization. 

The reconstructed spectra are shown in Figure~\ref{fig:all_cstars}. We can see that the spectra are all very similar, but do have small differences in the strength of the CN bands and the overall slope. Very few of the spectra have measurable C$_2$ at 7700\AA, just to the right edge of the telluric A-band masking. It is possible that this indicates that these spectra all have a fairly low C/O ratio, but it is more likely that the proximity of C$_2$ to the strong telluric feature has led to its being engulfed by the noise. Two of the spectra show prominent H$\alpha$ emission at $6562.8$\AA, and an additional two show weak emission. This fraction is far lower than the $\sim40\%$ of carbon stars observed with H$\alpha$ emission in the halo and satellites of the MW \citep{Mauron2007}.

\begin{figure}
\centering
\includegraphics[width=3.5in]{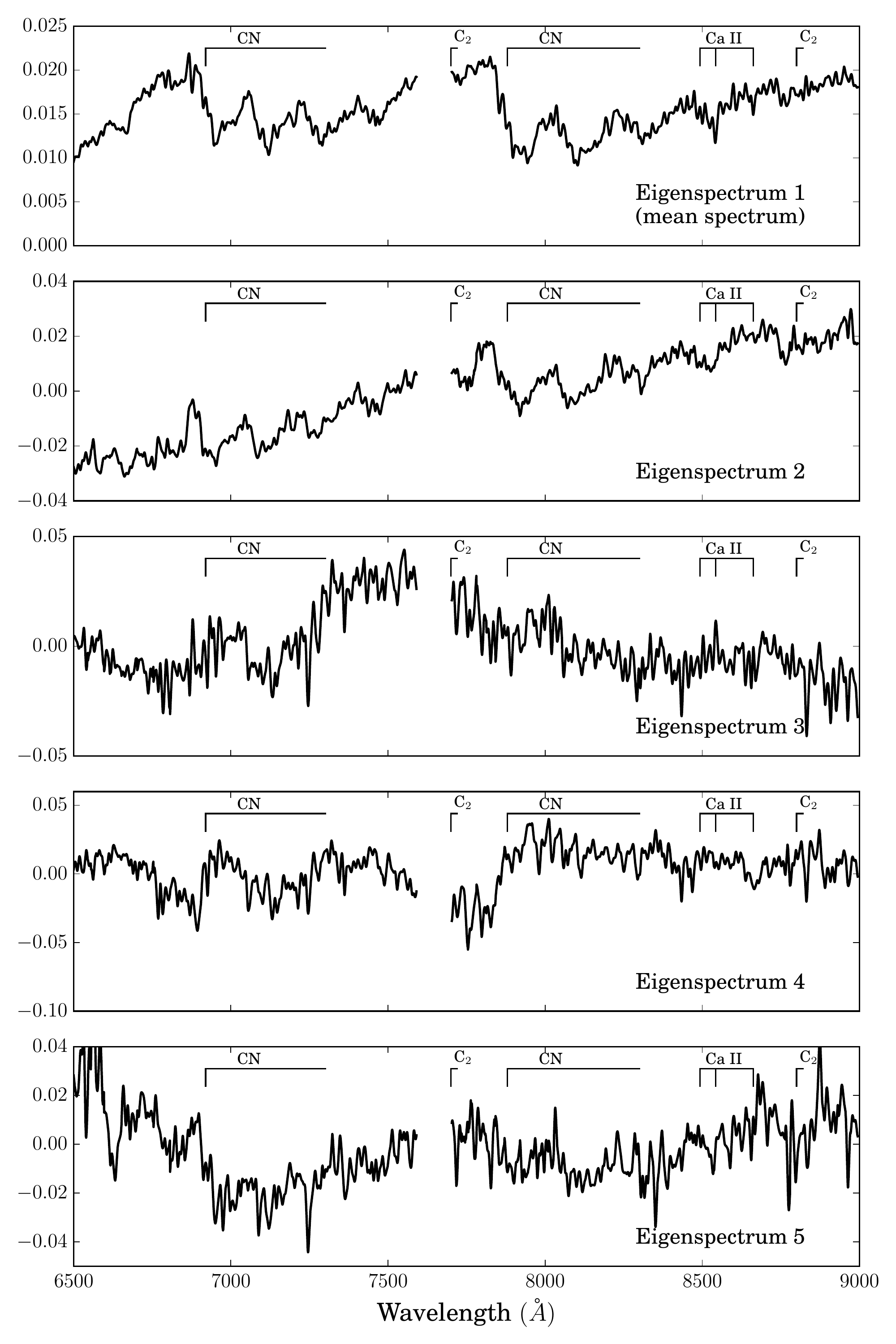}
\caption[First five C-star eigenspectra]{First five C-star eigenspectra. The eigenspectra are annotated with major absorption features (CN, C$_2$ and Ca {\sc II})}.
\label{fig:cstar_eigenspec}
\end{figure}

\begin{figure}
\centering
\includegraphics[width=3.5in]{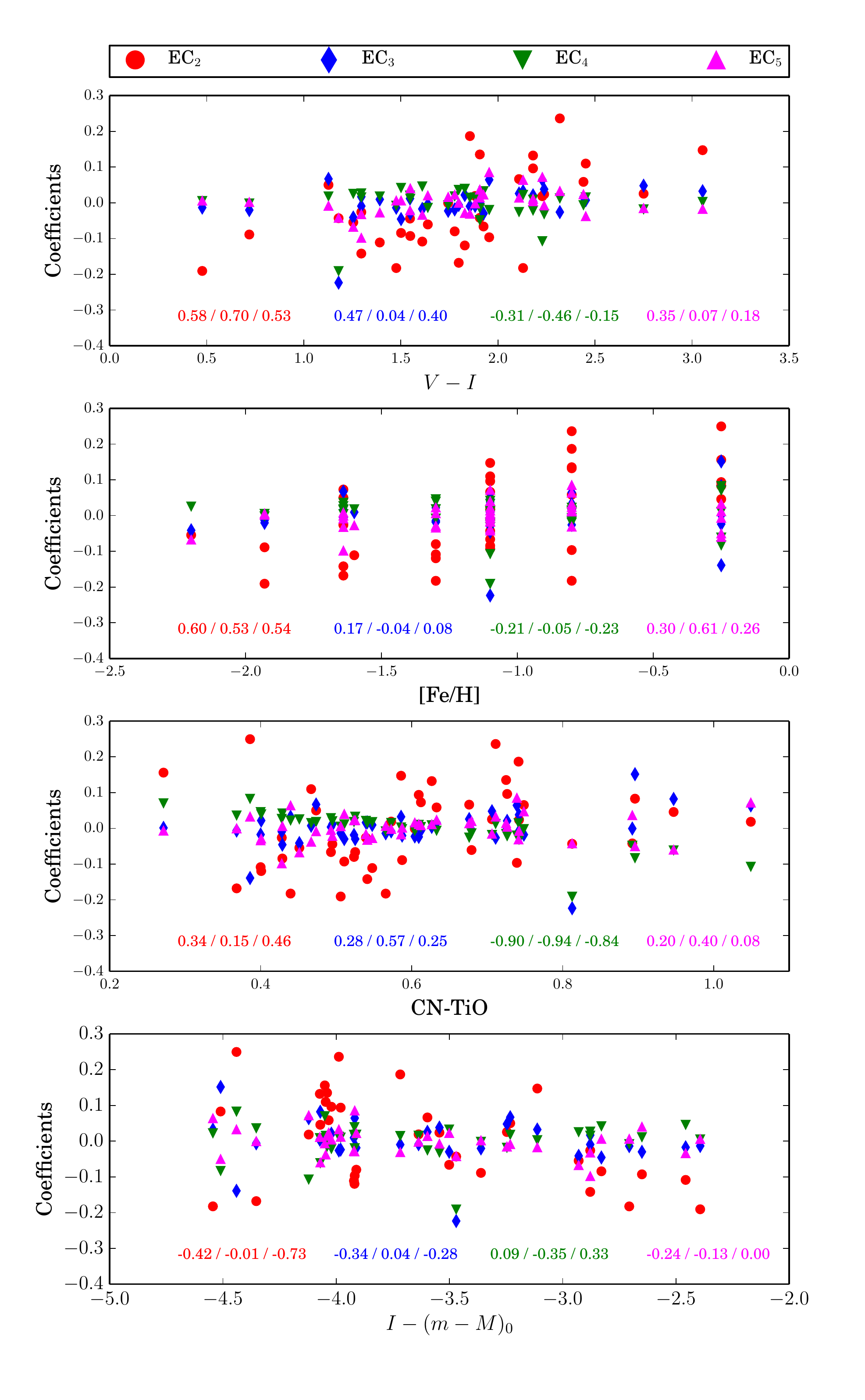}
\caption[C-star Eigencoefficient correlations]{C-star eigencoefficients versus physical properties ($V-I$ color, [Fe/H] of the host satellite, CN$-$TiO color, and absolute $I$-band magnitude) of the star. Coefficients of the second eigenspectrum are plotted as red circles, coefficients of the third are plotted as blue diamonds, coefficients of the fourth are plotted as inverted green triangles, and coefficients of the fifth are plotted as magenta triangles. The Spearman's rank correlation coefficient for each eigenspectrum is printed in the corresponding color, in the format ``all carbon stars / bright carbon stars only / faint carbon stars only." }
\label{fig:cstar_ecs}
\end{figure}

Figure~\ref{fig:cstar_eigenspec} shows the first five carbon star eigenspectra, which account for 42\% of the variance in the sample. The fifth eigenspectrum, as well as the subsequent eigenspectra not shown, is dominated by noise. Not only does it not correlate strongly with any of the physical properties shown in Figure~\ref{fig:cstar_ecs}, its eigencoefficients are all very small. With this noise, it takes 30 eigenspectra to account for 90\% of the variance in the carbon star sample. Higher S/N spectra will be necessary to tease out finer features than CN. 

The first eigenspectrum is the mean of the full carbon star sample. At a glance it is indistinguishable from many of the spectra in Figure~\ref{fig:all_cstars}, with prominent CN features. Visible here but not in each individual spectrum are some of the finer structure in the carbon features, such as the sawtooth shape of the CN band heads at $\sim6900$ and $\sim7900$\AA. There is also a visible Ca {\sc II} triplet, and what could be the C$_2$ feature at 7700\AA. However, there is no discernible signature of C$_2$ at 8800\AA.

The second eigenspectrum shows similar CN features to the first, as well as a general increase in flux at red wavelengths. Unlike the first eigenspectrum, it has no visible Ca {\sc II}, and stronger C$_2$ at 7700 and 8800\AA. This eigenspectrum governs, at least in part, the temperature and metallicity dependent aspects of the carbon star spectra. This suggests that the strength of CN varies with temperature and metallicity. The top two panels of Figure~\ref{fig:cstar_ecs} show the relationship between the eigencoefficients of this eigenspectrum (EC$_2$) as a function of $V-I$ color and [Fe/H] of the host satellite. The Spearman correlation coefficients (displayed on each panel in Figure~\ref{fig:cstar_ecs} for the full sample, the bright carbon stars alone, and the faint carbon stars alone, in that order) indicate positive correlations for both. This reflects the degeneracy seen in the models, where the combination of low metallicity and low temperature can result in a spectrum that looks remarkably similar at optical wavelengths to a star with high temperature and high metallicity \citep{Aringer2009, Aringer2016}. Interestingly, EC$_2$ also correlates with absolute $I$-band magnitude, but only for the faint (sub-TRGB) carbon stars.

The second eigenspectrum is also likely governed by the star's C/O ratio. At fixed temperature, metallicity, surface gravity, and mass, CN band strength increases with the C/O ratio \citep{Aringer2016}. However, its exact behavior depends on all of these properties as well as the star's nitrogen abundance. We are not able to reliably measure the C/O ratio in our sample, and so are unable to disentangle these competing effects.

\begin{figure}[t]
\centering
\includegraphics[width=3.5in]{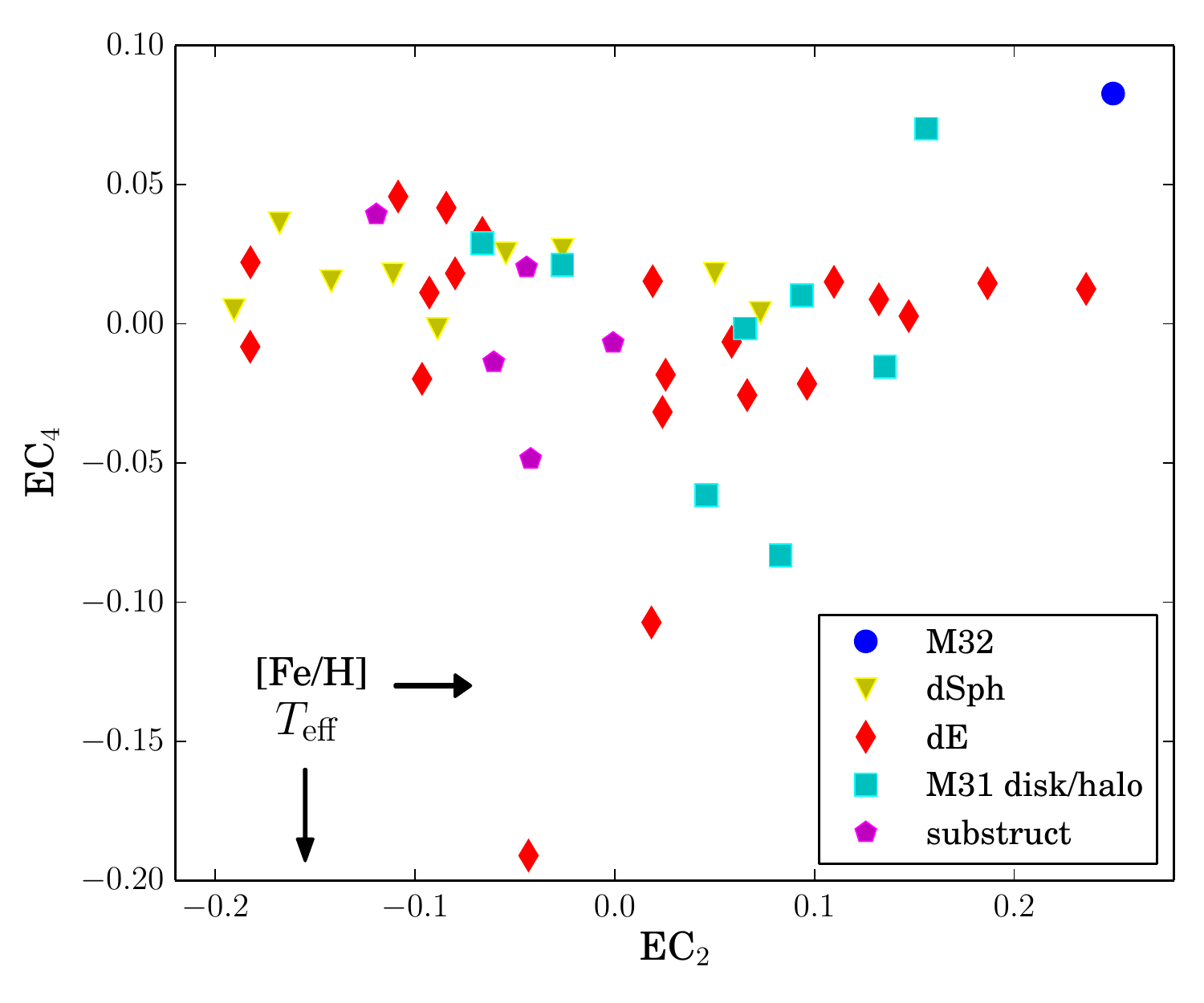}
\caption[EC$_2$ vs EC$_4$]{EC$_2$ (a tracer of metallicity) versus EC$_4$ (a tracer of $T_{\rm eff}$) for all carbon stars in the sample. Colors and shapes denote the type of field/object to which the star is a member, as designated in Table~\ref{tab:cstars}: stars in the dSphs are plotted as yellow triangles, stars in the dEs are plotted as red diamonds, stars in substructure fields are plotted as magenta pentagons, and the stars in M32 are plotted as blue circles. Finally, stars in the M31 disk or halo (which are kinematically difficult to separate) are plotted as cyan squares. }
\label{fig:ec2_ec3}
\end{figure}

The fourth eigenspectrum does not have the same CN features as the first two. Its most prominent feature is a break in flux at $\sim7900$\AA, just blueward of the 7900\AA~CN bandhead. There is a similar break just blueward of the 6900\AA~CN bandhead. Figure~\ref{fig:cstar_ecs} illustrates that it correlates strongly with CN$-$TiO color for the full carbon star sample as well as the two subsamples (with a Spearman correlation coefficient less than -0.8 for all three), but weakly with the other properties. Carbon star models indicate that at fixed metallicity CN$-$TiO color increases smoothly as $T_{\rm eff}$, whereas at fixed temperature the change in CN$-$TiO with metallicity is far less predictable \citep{Aringer2016}. The coefficients of this eigenspectrum (EC$_4$) are thus likely a more direct indicator of $T_{\rm eff}$ than EC$_2$, with EC$_2$ correlating inversely with $T_{\rm eff}$.

The third and fifth eigenspectra are more difficult to interpret. The only significant correlation seen in either comes out when looking only at the bright (super-TRGB) subset of carbon stars. The third eigenspectrum correlates with CN--TiO color, and the fifth eigenspectrum correlates with [Fe/H].

The eigencoefficients of the 2nd and 4th eigenspectra thus contain most of the information regarding the temperature and metallicity of the carbon stars. Figure~\ref{fig:ec2_ec3} shows EC$_4$ versus EC$_2$ for all carbon stars in the sample, coded by their environment (dE, dSph, M32, halo, or substructure). Following our interpretation of these eigencoefficients, this plot can be thought of as $T_{\rm eff}$ versus [Fe/H]. This figure suggests several trends. First, the single carbon star belonging to M32 has the highest values of both EC$_2$ and EC$_4$, indicating that it is both metal-rich and particularly cool. This is consistent with the metallicity of M32 being higher than the other satellites. Next, the carbon stars in the substructure and most dSphs fields have lower values of EC$_2$ than the other fields, but comparable values of EC$_4$. This suggests that they are more metal poor than the other fields. The dEs span the full range of EC$_2$.

The range of EC$_2$ values is far larger than the range of EC$_4$ values. While the eigencoefficient ranges do not correspond directly to the range in $T_{\rm eff}$ or metallicity, this may indicate that the carbon stars have comparable temperatures regardless of their metallicity. This is expected given the narrow mass/temperature range AGB stars occupy \citep{Suda2010, Karakas2014}.

\begin{figure}
\centering
\includegraphics[width=3.5in]{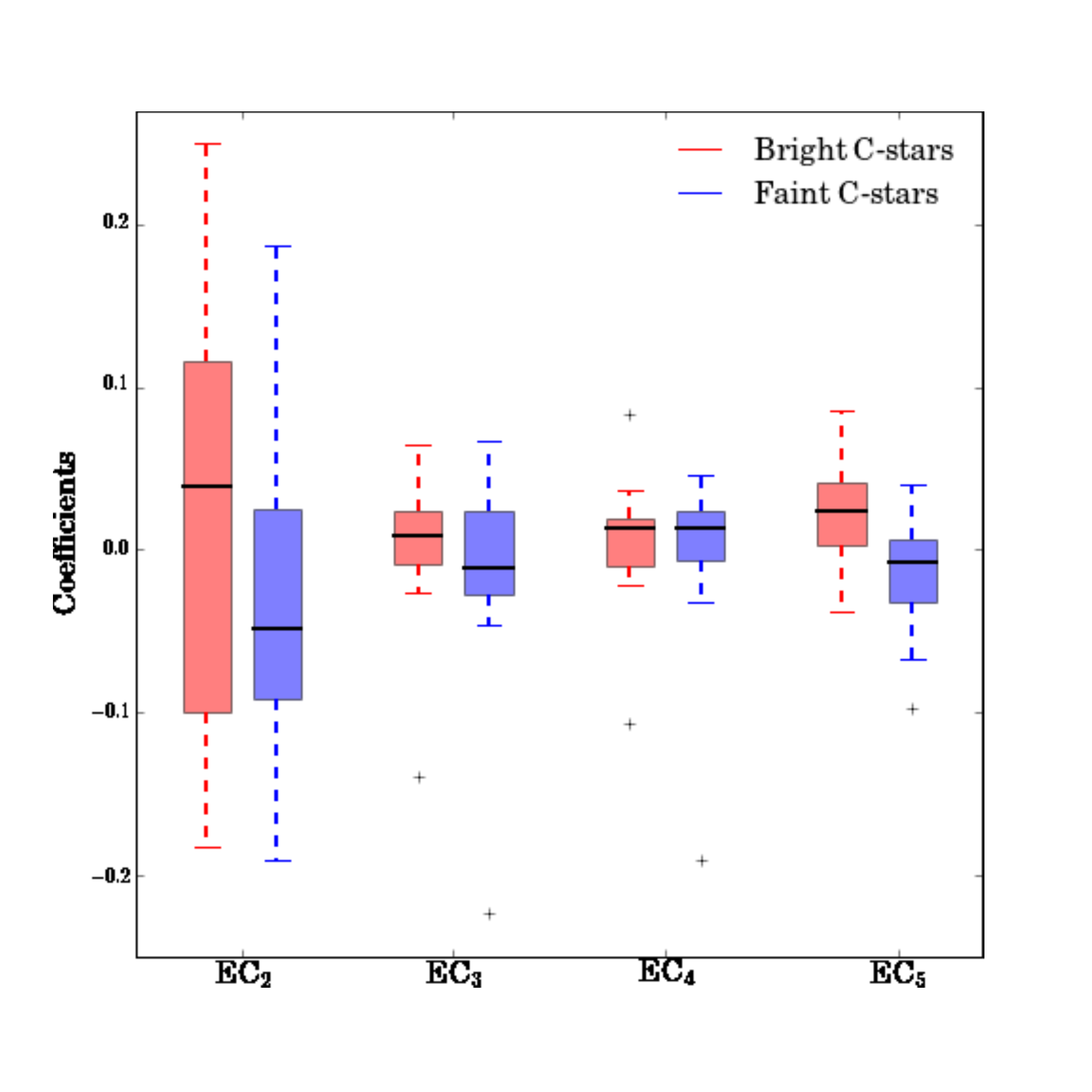}
\caption[ECs of bright vs faint carbon stars]{Eigencoefficients of bright (red) versus faint (blue) carbon stars. Black lines indicate the median of the distribution, red/blue boxes encompass the quartiles. Outliers are displayed as red/blue crosses.The distributions of eigencoeffients of the second and fifth eigencoefficients (EC$_2$ and EC$_5$) differ markedly, with the bright carbon stars typically having larger, more positive coefficients. The eigencoefficients of the third and fourth eigenspectra are nearly indistinguishable. }
\label{fig:bright_faint_ecs}
\end{figure}

None of the first five eigenspectra correlate strongly with absolute $I$ magnitude. There is thus no clear spectroscopic difference between the faint carbon stars in our sample and the bright carbon stars. Figure~\ref{fig:bright_faint_ecs} illustrates this more clearly, with boxplots representing the distribution of eigencoefficients for all bright and faint carbon stars. There is considerable overlap in the distributions of the faint and bright populations for each eigencoefficient. The two that show a significant difference are EC$_2$ and EC$_5$. The median EC$_2$ value for the bright carbon stars is higher than the majority of the faint carbon stars, which suggests that the bright carbon stars are more metal rich. 

\section{Discussion}\label{discussion}
In the previous sections we have presented the photometric and spectroscopic properties of the carbon stars found by the SPLASH survey in the satellites and halo of M31. This includes their distribution in various color-color spaces, their luminosities with respect to the $I_{\rm TRGB}$ of their respective regions, and the eigenspectra that make up the fundamental components of their spectra. Here, we will discuss the implications of these properties.

\subsection{General properties of the SPLASH carbon stars}

We identify 41 carbon stars in the satellites and halo of M31. Many are located in fields in which significant numbers of intermediate-age AGB stars have already been identified, including the dEs NGC~147, NGC~185 and NGC~205, and the dSph And~II \citep[e.g.][]{Nowotny2003, Kerschbaum2004, Battinelli2004a, Battinelli2004}. We also identify several carbon stars in the vicinity of M32, which is known to have a substantial intermediate-age population \citep{Davidge2014, Jones2015}. However, the majority of our sample have kinematics suggesting membership to M31 rather than M32 itself. We identify a small number of carbon stars in the dSphs And~V, And~IX and And~X. However this combined sample of four stars, three of which are fainter than the TRGB, indicates no significant intermediate-age population in these galaxies. Finally, we see four carbon stars in regions known to contain substructure from the Giant Southern Stream.

Analysis of the carbon star photometry also reveals that a significant fraction of our sample is fainter than the TRGB. These may be extrinsic in origin, and we will explore their nature fully in the next section. 

Our sample of carbon stars appears to be largely dust-free and hydrostatic. Their optical colors are fairly well fit by the dust-free hydrostatic models from \citet{Aringer2016}, though the models do not extend blue enough to fully match the colors of the faint carbon stars. The introduction of MIR colors complicates this picture, as the dust-free models do not match the observations well in optical versus MIR color-color space. However, none of the carbon stars in our sample appear in the DUSTiNGS variable star/extreme-AGB star catalog, which further supports the idea that they are not heavily effected by dust and dynamics (though again, two epochs is insufficient to rule out variability completely). Finally, very few of our carbon stars exhibit H$\alpha$ emission, which is associated with presence of shock waves induced by pulsation in the stellar atmosphere. Of the carbon stars in the solar neighborhood, 70\% of the Miras, 66\% of the SRa type, and 20\% of SRb and Lb type stars show H$\alpha$ in emission \citep{Mauron2014}. The carbon stars in the satellites and halo of M31 are comparatively quite quiet, which would be consistent with a significant population being extrinsic.

Application of PCA to the carbon star spectra illustrates the effects of temperature and metallicity. We find that the depth of the broad CN bands (governed by the second eigenspectrum) correlates most strongly with metallicity, while the depth and shape of the break in the spectrum at $\sim7900$\AA~traces $T_{eff}$ (governed by the fourth eigenspectrum). Comparing the coefficients of the eigenspectra governing these features illustrates that the carbon stars in M32 and the M31 disk/halo are typically more metal-rich than the carbon stars in the substructure and dSphs.  

\begin{deluxetable*}{lcccc}
\tablewidth{0pt}
\tablecaption{Fraction of faint carbon stars in the Local Group}
\tablehead{ Field & $N_C$ & $N_{FC}/N_C$ & $I_{\rm limit} - I_{\rm TRGB}$ & Reference}  \\
\startdata
IC 1613 & 195 & $0.087^{ 0.02}_{ 0.01}$ & 1.6 & \cite{Albert2000} \\
Leo I & 27 & $0.148^{ 0.06}_{ 0.04}$ & 1 & \cite{Demers2002} \\
Aquarius & 3 & $0.000^{ 0.16}_{ 0.00}$ & 0.85 & \cite{Battinelli2000} \\
Pegasus & 40 & $0.150^{ 0.05}_{ 0.04}$ & 1.1& \cite{Battinelli2000} \\
Sagittarius dIrr & 33 & $0.242^{ 0.06}_{ 0.05}$ &1.6 & \cite{Demers2002} \\
LMC & 7760 & $0.017^{ 0.00}_{ 0.00}$ & $\ldots$ & \cite{Kontizas2001} \\
NGC 6822 & 907 & $0.163^{ 0.01}_{ 0.01}$ & 1.9& \cite{Letarte2002} \\
Phoenix & 2 & $0.500^{ 0.23}_{ 0.23}$ & $\ldots$& \cite{Martinez-Delgado1999} \\
Draco & 6 & $1.000^{ 0.00}_{ 0.09}$ & $\ldots$& \cite{Shetrone2001} \\
Ursa Minor & 7 & $1.000^{ 0.00}_{ 0.07}$ &$\ldots$ & \cite{Shetrone2001} \\
And III & 1 & $1.000^{ 0.00}_{ 0.36}$ &1.6 & \cite{Harbeck2004} \\
And VI & 2 & $0.500^{ 0.23}_{ 0.23}$ & 1.6 & \cite{Harbeck2004} \\
And VII & 5 & $0.400^{ 0.17}_{ 0.15}$ & 1.4 & \cite{Harbeck2004} \\
Cetus & 3 & $0.667^{ 0.16}_{ 0.22}$ & 1.5& \cite{Harbeck2004} \\
And~X & 2 & $1.000^{ 0.00}_{ 0.22}$ & 1.92 & This work \\
M32 & 1 & $0.000^{ 0.36}_{ 0.00}$ & 0.25 & This work\\
And~IX & 1 & $1.000^{ 0.00}_{ 0.36}$ & 2.2 & This work\\
NGC 205 & 6 & $0.167^{ 0.14}_{ 0.08}$ & 0.3 & This work\\
And V & 1 & $0.000^{ 0.36}_{ 0.00}$ & 1.71& This work\\
NGC 147 & 158 & $0.301^{ 0.03}_{ 0.03}$ & 2.6& This work + \cite{Nowotny2003} \\
NGC 185 & 157 & $0.325^{ 0.03}_{ 0.03}$ & 3.0& This work + \cite{Nowotny2003} \\
And II & 10 & $0.300^{ 0.12}_{ 0.10}$ & 2.5 & This work + \cite{Kerschbaum2004}
\enddata
\label{tab:lit_cstars}
\end{deluxetable*}

\subsection{Characterizing the faint carbon stars}

In many of our satellites we see a significant number of carbon stars fainter than the TRGB. There are three possible explanations. The first is that these are extrinsic carbon stars, whose carbon was obtained by mass transfer from a carbon-rich AGB star onto a binary companion rather than through dredge-up of He-burning products during the TP-AGB phase \citep[][]{deKool1995, Frantsman1997, Izzard2004}. The second possible explanation is that they are genuine TP-AGB stars experiencing the post-flash luminosity dip \citep{Boothroyd1988b} or in a minimum of their dynamic phase \citep{Nowotny2011}. Finally, they may be bright AGB stars extincted by circumstellar dust. Both sections~\ref{photometry} and \ref{spectroscopy} treat the bright (super-TRGB) and faint (sub-TRGB) carbon stars as separate populations, to help determine which of the above explanations is most likely. 

From the photometry, we can rule out dust as a likely explanation for the faint carbon stars. If the faint carbon stars are heavily extincted then they should appear redder than the brighter population in $V-I$. However, as noted in the previous section, our sample appears to be well fit by the dust-free hydrostatic models. In addition, both the one and two dimensional color distributions (Figures~\ref{fig:boxplot} and \ref{fig:color_color}, respectively) indicate that the faint carbon stars are typically bluer in $V-I$ than the bright carbon stars. 

If the faint carbon stars are extrinsic, we would expect them to be present down to the SPLASH detection limit, while carbon stars on the TP-AGB would be clustered around the TRGB. The observed distribution of $I$-band magnitudes suggests the former: Figure~\ref{fig:below_trgb} illustrates that the faint carbon stars are present down to 1.5 magnitudes below the TRGB (though this varies from field to field given inhomogeneous detection limits). 

We would also expect extrinsic carbon stars to be warmer than their intrinsic counterparts. Both Figures~\ref{fig:boxplot} and \ref{fig:color_color} illustrate that many of the faint carbon stars are indeed bluer in $(V-I)$ than the bright carbon stars, despite comparable CN--TiO color. They are also typically bluer than the dust-free hydrostatic AGB models from \citet{Aringer2016}. The effects of dust and dynamics are not likely to make C-stars so optically blue ($V-I < 1.5$), so this suggests that these stars do not in fact belong to the AGB.

Spectroscopically, many of the known differences between intrinsic and extrinsic carbon stars fall beyond our wavelength coverage (e.g., G band of CH at $\sim4300$\AA, $^{12}C/^{13}C$ ratio). The major difference between the bright and faint carbon stars in our sample is the eigencoefficients of the second eigenspectrum (EC$_2$) (see Figure~\ref{fig:bright_faint_ecs}). The faint carbon stars typically have negative EC$_2$ values, while the bright carbon stars have positive EC$_2$ values. Observationally, this translates to the faint carbon stars having weaker CN bands and appearing more metal-poor than the bright carbon stars. This is fully consistent with observations of extrinsic carbon stars in the Local Group, which are often characterized by their lack of metals (e.g. the CEMP stars). Within the subset of faint carbon stars, we find that EC$_2$ anticorrelates with absolute $I$-band magnitude. Observationally, this means that the brightest/most massive of the faint carbon stars have the strongest CN. Finally, the eigencoefficients of the fifth eigenspectrum (EC$_5$) also differ between the bright and faint samples, perhaps due to more noise in the fainter stars. 

However, all of the figures discussed in this section also illustrate that the colors and magnitudes of the faint carbon stars cover a significant range. Some are fully consistent with the bright carbon stars. It is probable that the faint population consists of a mix of intrinsic and extrinsic carbon stars. Models based on the Magellanic Clouds suggest that $<10\%$ of TP-AGB stars are fainter than the TRGB \citep[e.g.,][]{Marigo2007, Melbourne2013} at any given time. This places an upper limit on the number of intrinsic interlopers in the faint population at 1.8 stars. 

\subsection{Carbon star luminosity by environment}

\begin{figure*}
\centering
\includegraphics[width=7in]{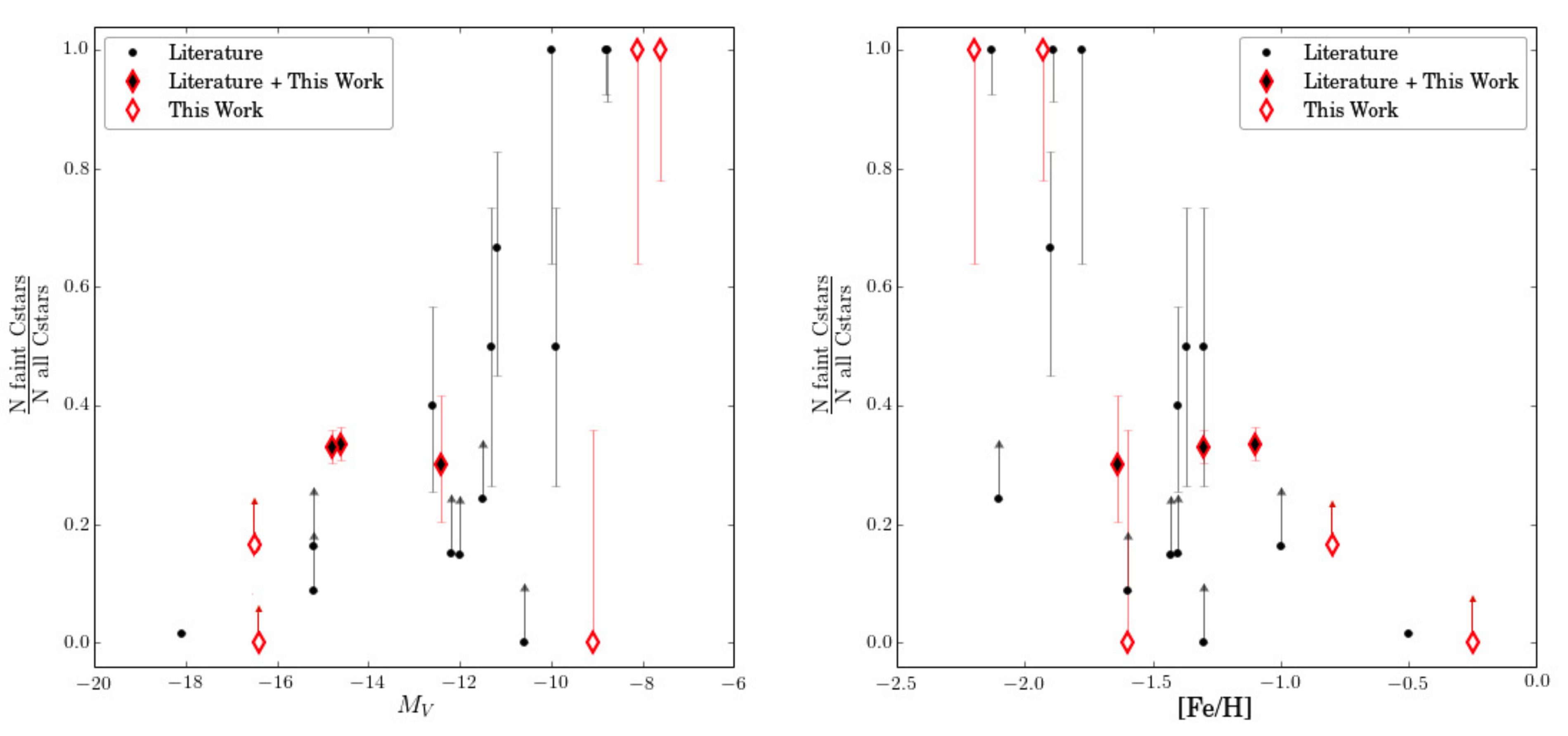}
\caption[Fraction of faint carbon stars by environment]{Fraction of faint carbon stars as a function of [Fe/H] of the host satellite. Samples from the literature are plotted as black points, samples from the present work are plotted as red diamonds, and samples that combine new and existing data are plotted as red diamonds filled with black. Also shown are the 1$\sigma$ binomial proportion confidence intervals. Though the uncertainties are large, the fraction of the full carbon star sample fainter than the TRGB appears to increase in smaller, metal-poor galaxies. }
\label{fig:lit_cstars}
\end{figure*}

At first glance, we appear to observe more faint carbon stars in our less luminous galaxies. To parametrize this, we calculate $N_{FC}/N_{C}$, the fraction of faint (sub-TRGB) carbon stars in the full sample of carbon stars, for each galaxy. To place these findings within a broader context, we also calculate $N_{FC}/N_{C}$ in other Local Group satellites. To compare similar groups of stars, we limit ourselves to optical carbon star surveys, leaving aside surveys in the NIR and MIR and serendipitous carbon (or CH) star discoveries. We assemble C-star counts in IC~1613 \citep{Albert2000}, Leo~I, Sagittarius dIrr \citep{Demers2002}, DDO~210/Aquarius, Pegasus \citep{Battinelli2000}, NGC~6822 \citep{Letarte2002}, Phoenix \citep{Martinez-Delgado1999}, Draco, Ursa Minor \citep{Shetrone2001}, the LMC \citep{Kontizas2001}, And~III, And~VI, And~VII and Cetus \citep{Harbeck2004}. These data are provided in Table~\ref{tab:lit_cstars}.

The majority of these fields (IC~1613, Pegasus, Aquarius, NGC~6822, Leo I, Sag dIrr) rely on optical FBPS to identify carbon stars. All survey the full galaxy area and extend to 1-2 magnitudes fainter than the TRGB. The approximate extent below the TRGB is listed in Table~\ref{tab:lit_cstars}. This is roughly equivalent to the SPLASH coverage and detection limit, so we will not worry tremendously about differences in depth for these fields. However, all require C-stars to have $R-I > 0.9$ (though \citealt{Letarte2002} also look at ``bluer" C-stars with $0.8 < R-I < 1.1$). This makes them biased against the faint carbon stars, which we find to be bluer. As a result, the $N_{FC}/N_C$ that we calculate in these fields are likely lower limits. In each field we compute $I_{\rm TRGB}$ following the same prescription outlined in Section~\ref{selection functions} and count the number of C-stars fainter than that limit.

We consider the carbon star samples in Phoenix, Draco, Ursa Minor, the LMC, And~III, And~VI, And~VII and Cetus to be representative of the full carbon star population. The two carbon stars in Phoenix were confirmed spectroscopically by \citet{DaCosta1994}, though our data come from \citet{Martinez-Delgado1999}. Subsequent work by \citet{Menzies2007} did not identify any other carbon stars visible in the optical, so we consider this sample of two to be complete. The carbon star census presented in \citet{Shetrone2001} for Draco and Ursa Minor relies on the identification of carbon stars via photographic plates \citep{Aaronson1982, Azzopardi1986}. These stars have been thoroughly vetted \citep[e.g.,][]{Dominguez2004,Abia2008}, and no new carbon stars have been identified. The Kontizas et al. survey of the LMC was conducted with objective-prism plates and visual identification of carbon-stars using the Swan C$_2$ bands at 4737\AA~and 5165\AA. Because this is a spectroscopic rather than photometric identification, there is not a strong bias against faint C-stars. This survey is likely incomplete in the most crowded central regions, but we do not expect the extrinsic and intrinsic carbon stars to have different spatial distributions and this incompleteness is not likely to skew our counts. Thus, the $N_{FC}/N_{C}$ fraction we calculate in the LMC is likely representative. These four fields rely on spectroscopic identification of carbon stars, so their photometric limits are excluded from Table~\ref{tab:lit_cstars}. Finally, \citet{Harbeck2004} use FBPS to identify carbon stars in And~III, And~VI, And~VII and Cetus, but do not apply a limit in $V-I$. They thus sample the bright and faint carbon stars uniformly.

Because \citet{Shetrone2001} provide $V$-band magnitudes of the carbon stars in Draco and Ursa Minor, we compute $V_{\rm TRGB}$ using relationships from \citet{Bellazzini2004} and \citet{Mager2008}. For Phoenix and the LMC, for which provide $I$-band magnitudes have been provided, we once again calculate the TRGB magnitude using the method described in Section~\ref{selection functions}. \citet{Harbeck2004} have provided designations of C versus dC (which translate to super- and sub-TRGB), which we use to calculate $N_{FC}/N_{C}$ in their four fields.

In addition to the new fields discussed above, we also compile carbon star populations in And~II, NGC~147 and NGC~185 from the literature. We combine our carbon star samples observed by SPLASH with these samples from the literature. The observations of And~II \citep{Kerschbaum2004}, NGC~147 and NGC~185 \citep{Nowotny2003} also use FBPS, but with $V$ and $i_0$ as the broad-band colors rather than $R$ and $I$. Both authors' observations extend $\sim0.5$ magnitudes below the TRGB, and they require C-stars to have $(V-i)_0 > 1.16$ and (TiO--CN)$_0 < -0.3$. This limit in $(V-i)_0$ is equivalent to $(R-I) > 0.53$, which encompasses the full color range over which we observe carbon stars. As a result, we consider this a representative sample. As \citet{Bellazzini2004} do not provide a TRGB calibration for $i_0$, we use the values calculated by \citet{Kerschbaum2004}, $i_{\rm TRGB} = 20.5$, and \citet{Nowotny2003}, $i_{\rm TRGB} = 19.96$ for NGC~185 and $i_{\rm TRGB} = 20.36$ for NGC~147, when computing N$_{FC}/N_C$.

Finally, two of the SPLASH fields (M32 and NGC~205) were surveyed so shallowly that they too should be considered lower limits on N$_{FC}/N_C$. We calculate this limit in SPLASH fields using the full SPLASH sample (independent of membership or strength of carbon features).

Figure~\ref{fig:lit_cstars} plots $N_{FC}/N_{C}$ in the galaxies listed in Table~\ref{tab:lit_cstars} as a function the absolute magnitude and metallicity of the host satellite \citep[obtained from][]{McConnachie2012}. Also plotted are the $1\sigma$ binomial proportion confidence intervals, save in the cases discussed above where the calculated fraction represents a lower limit. With the exception of the LMC, NGC~147 and NGC~185, the small sample sizes translate to large uncertainties. A trend is still evident; the fraction of faint carbon stars decreases as the galaxy gets brighter and its metallicity increases. The serendipitous detections of CH and dC stars in MW globular clusters adheres to this trend, increasing the number of faint, low-metallicity systems in which $N_{FC}/N_{C} = 1$.

There are two possible explanations for this trend. The first presumes that the faint carbon stars are extrinsic and owe their composition to a binary companion. If the fraction of stars in binary systems increases in smaller satellites \citep[as it has been shown to increase in smaller globular clusters,][]{Milone2012}, then so will the fraction of faint carbon stars. However this explanation would also require a sizeable population of AGB stars in earlier generations to provide the binary companions. The second explanation presumes that the entire carbon star sample is made up of intrinsic TP-AGB stars. In this case, the trend could be explained by the effect of metallicity on carbon star formation. At low metallicity, dredge-up efficiency is higher, the range of masses over which carbon stars can form is larger, and TDU begins earlier \citep{Karakas2002, Marigo2013, Karakas2014}. It is possible that at low metallicities these effects combine to produce carbon stars that are fainter than the TRGB. Indeed both explanations may be occur simultaneously. The more massive companion in an earlier binary pair is likely to go through the AGB. In metal-poor galaxies, that AGB star is more likely to be carbon-rich and the low-mass companion is more likely to become carbon enhanced. 

\section{Conclusions}

We identify 41 unambiguous carbon stars in the satellites and halo of M31. We present optical, synthetic narrow-band, and MIR photometry of these stars, as well as moderate-resolution optical spectra. Photometric and spectroscopic analysis suggests that they are relatively unaffected by dust and dynamics.

Many of the carbon stars we identify are fainter than the TRGB. In addition to being fainter, these stars are also often bluer and more metal-poor than their super-TRGB counterparts. They are likely to be extrinsic carbon stars. However, this designation is far from unambiguous. Observations at different wavelengths (bluer to capture features known to distinguish intrinsic and extrinsic chemistry, redder to capture any effects of dust), and the extension of carbon-star models to low metallicity ([Fe/H]$< -1.5$) will be necessary to classify these stars definitively.

\acknowledgements
The authors would like to thank Bernhard Aringer and L\'{e}o Girardi for helpful conversation and an early look at the 2016 cool star models. We would also like to thank Marla Geha, James Bullock and Jason Kalirai for their work on the SPLASH survey over the years, and their willingness to provide data for this paper. PG and KH acknowledge NSF grants AST-1010039 and AST-1412648 and NASA grant HST-GO-12055. RLB and SRM thank NSF grant AST-1413269. KH was supported by a NSF Graduate Research Fellowship and EJT was supported by a Giacconi Fellowship. We appreciate the very significant cultural role and reverence that the summit of Mauna Kea has always held within the indigenous Hawaiian community. We are most grateful to have had the opportunity to conduct observations from this mountain.

\bibliographystyle{apj}
\bibliography{all_references}

\end{document}